\documentclass{article}
\usepackage{graphicx} 
\usepackage{amsmath}
\usepackage{amsfonts}
\usepackage{fixmath}
\usepackage{multirow}
\usepackage{bm}
\usepackage{color}
\usepackage{url}
\usepackage[numbers]{natbib}
\usepackage{appendix}
\usepackage{setspace}
\usepackage{lineno}
\usepackage{enumitem}
\title{Fast spatio-temporally varying coefficient modeling with reluctant interaction selection}

\author{Daisuke Murakami$^{1,2,\ast}$, Shinichiro Shirota$^{3}$, Seiji Kajita$^{1}$, Mami Kajita$^{1}$ }

\begin{document}
\date{}
\maketitle
\begin{center}

\textit{
$^{1}$ Singular Perturbations Inc, Japan; e-mail: dmuraka@ism.ac.jp\\
$^{2}$ Institute of Statistical Mathematics, Japan \\
$^{3}$ Hitotsubashi University, Japan \\
$\ast$ Corresponding author}
\end{center}
\vskip\baselineskip
\begin{abstract}
Spatially and temporally varying coefficient (STVC) models are attracting attention as a flexible tool to explore the spatio-temporal patterns in regression coefficients. However, these models often struggle with balancing computational efficiency, flexibility, and interpretability of the coefficients. To address this challenge, this study develops a fast and flexible STVC model. To enhance flexibility and interpretability, we assume multiple processes in each varying coefficient, including purely spatial, purely temporal, and spatio-temporal interaction processes with or without time cyclicity. We combine a pre-conditioning method with a model selection procedure, inspired by reluctant interaction modeling, to estimate the strength of each process in each coefficient in a computationally efficient manner, while removing redundant processes as necessary. Monte Carlo experiments demonstrate that the proposed method outperforms alternatives in terms of coefficient estimation accuracy and computational efficiency. We then apply the proposed method to a crime analysis. The result confirms that the proposed method provides reasonable estimates. The STVC model is implemented in an R package spmoran.
\end{abstract}

\section{Introduction}\label{sec_intro}
Spatio-temporally varying coefficient (STVC) models have been used to explore the relationship between explained and explanatory variables varying over space and time. Geographically and temporally weighted regression (GTWR; \cite{huang2010geographically,fotheringham2015geographical}) is the most popular STVC model, while the Gaussian process \cite{williams1995gaussian} and generalized additive model (GAM; \cite{wood2017generalized}) have also been used for estimating/predicting STVCs. These models have been applied to analyze crime occurrence \cite{murakami2020scalable}, carbon emission \cite{he2023spatial}, habitat quality \cite{hu2022exploring}, and epidemic \cite{sifriyani2022spatial}.

\begin{figure}
    \centering
    \includegraphics[width=11.5cm]{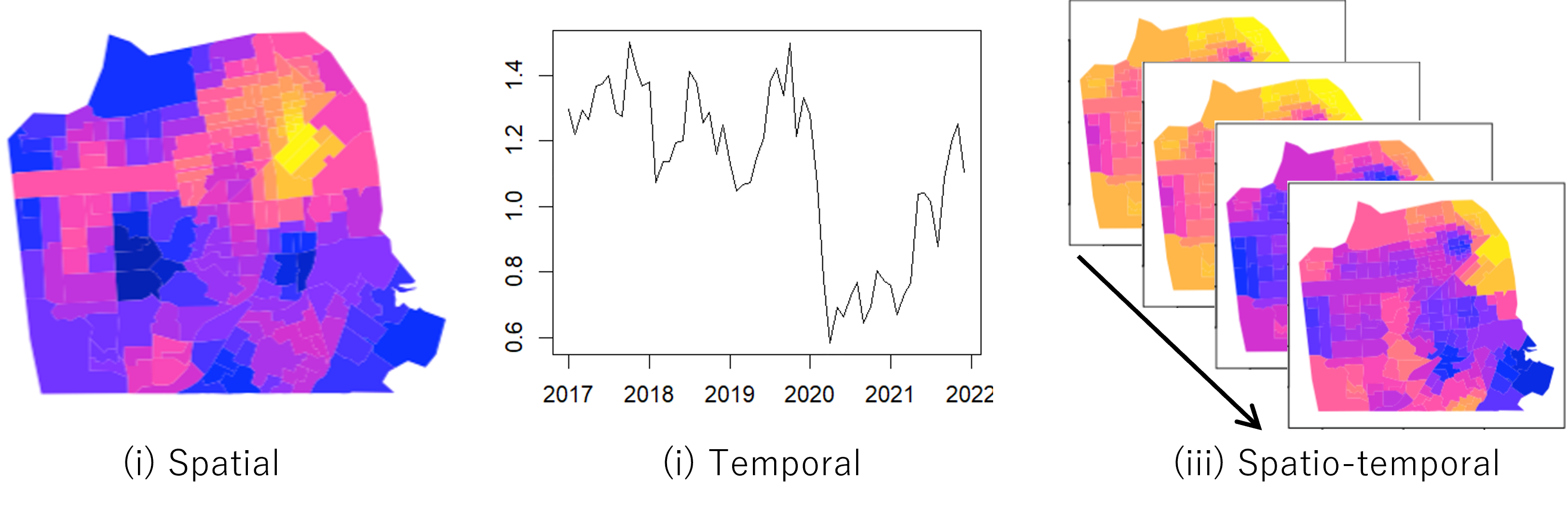}
    \caption{(i) Spatial, (ii) temporal, and (iii) spatio-temporal patterns.}
    \label{Fig_i_iii}
\end{figure}

Although each STVC is typically assumed to obey a single spatio-temporal process, it may depend on multiple processes. Specifically, regression coefficients can exhibit (i) spatial, (ii) temporal, and/or (iii) spatio-temporal variations (Fig. \ref{Fig_i_iii}). For example, when modeling crime density, the regression coefficients may depend on multiple processes, including a process spatially varying depending on neighborhood characteristics (e.g., income level), a process varying temporally depending on macroeconomic conditions, and a process varying spatio-temporally in response to urban activities over time. To enhance the interpretability and stability of the STVC modeling, it will be useful to estimate the proportion of (i – iii) in each coefficient while eliminating redundant processes, if necessary.

In addition, (ii) and (iii) can exhibit either non-cyclic or cyclic temporal variation. For instance, in the crime example, coefficients may vary non-cyclically depending on urban expansion over the years, while they may vary cyclically throughout a day, week, and year. Unfortunately, if all the aforementioned patterns are considered, the resulting STVC model is highly complex, leading to an overfitting and a high computational burden.

Therefore, this study develops a fast and flexible STVC modeling approach that involves constant coefficients, spatially varying coefficients (SVC), cyclic/non-cyclic temporally varying coefficients, and cyclic/non-cyclic STVC, corresponding to (i), (ii), and (iii), respectively. To improve stability and computational efficiency, we propose a practical model selection method inspired by the idea of reluctant interaction modeling \cite{yu2019reluctant,tay2020reluctant}. Based on the principle that "one should prefer main (or non-interaction) effects over interactions if all else is equal," this approach estimates main/non-interaction processes first, followed by interaction processes. In our context, (i) and (ii) correspond to the main/non-interaction processes and (iii), which models interactions between space and time, corresponds to the interaction process.

The remainder of the sections are organized as follows. Section \ref{sec_review} reviews related models. Section \ref{sec_method} explains our proposed method. Sections \ref{sec_mc} and \ref{sec_app1} assess the modeling accuracy and computational efficiency of our method through Monte Carlo experiments and an empirical application, respectively. Section \ref{sec_remark} concludes our discussion. 

\section{Related models}\label{sec_review}
This section reviews related models, including GTWR, GP, and GAMs, for spatio-temporal data. As explained below, the challenges associated with these models are as follows: GTWR: inflexibility and high computational cost; GP: high computational cost; GAM: lack of accuracy due to a degeneracy problem. The developed STVC model to overcome these challenges is described in Section \ref{sec_method}.

\subsection{Geographically and temporally weighted models}\label{sec_review_gwr}
Conventional GTWR estimates coefficients at site $s \in \mathbb{R}^2$ at time $t \in \mathbb{R}^1$ through a least squares estimation using sample weights based on the space-time kernel $k_{h_s}(d_{s,s'})k_{h_t}(d_{t,t'})$. $d_{s,s'}$ represents the Euclidean distance between locations $s$ and $s'$, $d_{t,t'}$ represents the time difference between $t$ and $t'$, and $h_s$ is the spatial bandwidth parameter; a larger $h_s$ yields a larger-scale map pattern in the coefficients. $h_t$ is another bandwidth that determines the temporal scale.

As explained in Section \ref{sec_intro}, distinguishing (i) spatial, (ii) temporal, and (iii) space-time interaction is important to improve the interpretability and flexibility of the coefficients. In this regard, the product kernel $k_{h_s}(d_{s,s'})k_{h_t}(d_{t,t'})$ ignores (i) and (ii). Although GTWR has been extended for flexible modeling \cite{
wu2014geographically,liu2017mixed,wu2019multiscale,wu2021geographically,li2022enhanced}, to the best of the author's knowledge, (i) - (iii) have not been considered simultaneously in GTWR studies. Considering the computational cost of optimizing the bandwidth parameters \cite{liu2017mixed}, extending GTWR to accommodate (i - iii), which requires additional bandwidth parameters, may be less reasonable. GTWR faces challenges in flexibility and computational efficiency.

\subsection{Gaussian process models}\label{sec_review_gp}

While GTWR weights samples, GP parameterizes covariance using a kernel $k_{\bm{\theta}}(d_{s,s'},d_{t,t'})$ depending on parameters $\bm{\theta}$ to model a spatio-temporal process. As reviewed in \cite{porcu202130}, various kernels have been proposed in geostatistics. For example, $k_{\bm{\theta}}(d_{s,s'},d_{t,t'})=k_{\bm{\theta}}(d_{s,s'})k_{\bm{\theta}}(d_{t,t'})$, which is also used in GTWR studies, and $k_{\bm{\theta}}(d_{s,s'},d_{t,t'})=k_{\bm{\theta}}(d_{s,s'})+k_{\bm{\theta}}(d_{t,t'})$ are recognized as basic specifications in the literature. Among more sophisticated kernels (e.g., \cite{gneiting2002nonseparable, de2001product,cressie1999classes}), the product-sum kernel \cite{de2001product} is popular due to its simplicity and flexibility. This kernel is defined as $k_{\bm{\theta}}(d_{s,s'},d_{t,t'})=k_{\bm{\theta}}(d_{s,s'})+k_{\bm{\theta}}(d_{t,t'})+\xi k_{\bm{\theta}}(d_{s,s'})k_{\bm{\theta}}(d_{t,t'})$  where $\xi$ is a parameter. The three terms in this kernel characterize (i) spatial, (ii) temporal, and (iii) spatio-temporal processes, respectively.

While the above-mentioned kernels assume non-cyclic temporal patterns, numerous studies have developed kernels for modeling space and circular time processes (e.g., \cite{ShirotaGelfand(17a),JonaGelfandJona(12),WangGelfand(14)})\footnote{Wrapping time to a circle takes us to the realm of directional or angular data where we find applications to environmental sciences, e.g., wind directions and wave directions. Traditional models for directional data have employed the von Mises distribution; however, recent work has elaborated the virtues of the wrapping and projection approaches, particularly in space and space–time settings (see \cite{JonaGelfandJona(12)} and \cite{WangGelfand(14)}). For a review of directional data methodology, see, for example, \cite{Fisher(93)}, \cite{Mardia(72)}, and \cite{MardiaJupp(00)}.}. These cyclic kernels capture the overnight dependence such that 23:55 and 00:05 are as close as 23:45 and 23:55. Cyclic kernels are also helpful for modeling temporal processes over a week and a year.

A challenge with GP is the computation burden arising from the inversion of the $N \times N$ covariance kernel matrix, where $N$ is the sample size. This burden is even more severe for the GP-based S(T)VC model, as it requires a heavy inversion for each varying coefficient (see \cite{gelfand2003spatial}).

To lighten the computation cost, a wide variety of approximate GPs have been proposed in geostatistics, as reviewed in \cite{heaton2019case} and \cite{hazra2024exploring}. Existing methods approximate GPs using $L (<<N)$  basis functions (see \cite{cressie2022basis} for review), sparse covariance matrix (e.g., \cite{furrer2006covariance}), or sparse precision matrix (e.g., \cite{lindgren2011explicit}, \cite{datta2016hierarchical} ). The nearest neighbor Gaussian process (NNGP; \cite{datta2016hierarchical}, \cite{datta2016nearest}), which is categorized in the third approach, is among popular approaches whose accuracy and scalability have been demonstrated. Especially, the conjugate NNGP \cite{finley2019efficient}, which employs a cross-validation instead of the Markov-Chain Monte Carlo simulation that makes the original NNGP slow, is extremely fast while maintaining the accuracy. However, as noted by the authors, the cross-validation can be inefficient in richly parameterized settings like our case, assuming variance parameters for each space, time, and space-time process by coefficients (Section \ref{sec_method}). Unfortunately,  most approximate GPs do not assume such highly parameterized settings. It is unclear how to apply them for modeling our STVCs. 

Exceptionally, GAM, which is a basis function approach that can approximate geostatistical GP \cite{kammann2003geoadditive}, has been extended to adapt to such highly parameterized settings. The next subsection reviews studies on GAM.

\subsection{Generalized additive models}\label{sec_review_gam}

GAM accommodates various patterns using a linear combination of basis functions \cite{wood2017book}. Owing to its parsimonious model structure and fast estimation algorithms (e.g., \cite{wood2011fast}), GAM is computationally more efficient than GTWR and GP models. 

When modeling STVCs, GAM uses $L$ spatial and $L_t$ temporal basis functions for modeling (i) and (ii), respectively. For (iii), GAM typically considers $LL_t$ basis functions defined by their products. If $L$ and/or $L_t$ are too small, the resulting
varying coefficient becomes over-smoothed and less accurate. This degeneracy problem \cite{stein2014limitations} becomes severe when considering (iii) because a large number of basis functions are needed to accurately model space-time interactions. Unfortunately, the large basis dimension makes GAM slow. Therefore, extending GAM to accommodate a large number of basis functions while maintaining its computational efficiency remains an important issue.

\section{Methodology} \label{sec_method}

We develop a fast and flexible STVC model that defines each varying coefficient as a sum of constant (mean), (i) spatial, (ii) $Q$ cyclic/non-cyclic temporal, and (iii) $Q$ cyclic/non-cyclic spatio-temporal processes, each of which are defined by a linear combination of basis functions. Section \ref{subsec_model} introduces our STVC model and Section \ref{subsec_select} describes its estimation and selection procedure.

\subsection{Model}\label{subsec_model}
This study considers the explained variable $y(s,\bm{t})$ observed at site $s \in \{1, \ldots, S\}$ in a study region $D \subset \mathbb{R}^2$  at time $\bm{t}=\{t_1,\ldots,t_Q\}$ measured on several axes indexed by $q \in \{1,\ldots, Q\}$. For example, $q=1$ and $q=2$ represent month and hour in the empirical study in Section \ref{sec_app1}, respectively. We assume a longitudinal data with/without missing values, whose explained variables are observed repeatedly at the S sample locations.

\subsubsection{Model for data}\label{subsec_process}
We consider the following model:
\begin{equation}
y(s,\bm{t})=\sum_{p=1}^P x_p(s,\bm{t}) \beta_p(s,\bm{t}) + \epsilon(s,\bm{t}),\hspace{0.5cm}
\epsilon(s,\bm{t}) \sim N(0,\sigma^2),   \label{eq_data}
\end{equation}
where $x_p(s,\bm{t})$ is the $p$-th covariate, with $p=1$ assumed to be constant (i.e., $x_1(s,\bm{t})=1$), and $\sigma^2$ is the noise variance. The coefficient $\beta_p(s,\bm{t})$ is specified as
\begin{equation}
\beta_p(s,\bm{t})=b_p+w_p(s)+\sum_{q=1}^Q w_p(t_q) +\sum_{q=1}^Q w_p(s,t_q),\label{eq_beta}
\end{equation}
where $b_p$ denotes the mean. $w_p(s),w_p(t_q)$, and $w_p(s,t_q)$ are zero-mean processes explaining (i), (ii), and (iii), respectively. $\beta_p(s,\bm{t})$ represents a varying intercept for $p=1$ while varying coefficients for $p \ge 2$. Instead of the independent noise $\epsilon(s, \bm{t})$, the varying intercept captures residual spatio-temporal dependence (see Section \ref{subsec_process}).

\subsubsection{Model for processes}\label{subsec_process}
To specify the spatial process $\omega_p(s)$, we consider the Moran's I statistic, which measures the strength of spatial dependence of a vector $\bm{\omega}_p$ whose elements are $\omega_p(s)|s \in \{1,\ldots,S\}$, as follows:
\begin{equation}
I[\bm{\omega}_p]=c\frac{\bm{\omega}_p\bm{MKM}\bm{\omega}'_p } {\bm{\omega}_p\bm{M}\bm{\omega}'_p}.
\end{equation}\label{eq_mi}
$c=S/\bm{1K1}'$ is a scaling factor. $\bm{M}=\bm{I} - \bm{1}\bm{1}' /S$ is a centering matrix where $\bm{I}$ is an identity matrix, and $\bm{1}$ is a vector of ones. $\bm{K}$ is a spatial kernel matrix with zero diagonals. This statistic indicates $I[\bm{\omega}_p]>0$ in the presence of positive spatial dependence, whereas $I[\bm{\omega}_p] \approx 0$ in the absence of spatial dependence.

Let us eigen-decompose $\bm{M}\bm{K}\bm{M}$ into $\bm{E} \bm{\Lambda} \bm{E}'$, where $\bm{\Lambda}$ is a diagonal matrix whose elements are the positive eigenvalues $\lambda_1,\ldots,\lambda_L$ in descending order and $\bm{E}=[\bm{e}_1,\ldots, \bm{e}_L]$ consists of the corresponding eigenvectors ($L<N$). \cite{griffith2003spatial} proved that $I[\bm{e}_l]=c\lambda_l$, meaning that $I[\bm{e}_1] \ge I[\bm{e}_2], \ldots, \ge I[\bm{e}_L]$, where $I[\bm{e}_1]=c\lambda_1$ is the theoretical maximum of the Moran's I statistic and $I[\bm{e}_L]=c\lambda_L \approx 0$. In other words, the $L$ eigenvectors facilitate as basis functions describing positively dependent map patterns in each level.

The spatial process $\omega_p(s)$ is specified as follows:
\begin{equation}
w_p(s)= \bm{e}(s) \bm{\gamma}_{p,0}, \hspace{1cm} \bm{\gamma}_{p,0} \sim N(\bm{0},  \tau_{p,0}^2 \bm{\Lambda}^{\alpha_{p,0}} ),\label{eq_w_s}
\end{equation}
where $\bm{e}(s)$ is the $s$-th row of the $\bm{E}$ matrix.  $\tau_{p,0}^2$ and $\alpha_{p,0}$ determine the variance and smoothness of the process. The latter also determines the strength of the spatial dependence. $I[\bm{\omega}_p]$ approaches the theoretical maximum, suggesting marked spatial dependence when $\alpha_{p,0} \to \infty$, while approaches zero, suggesting the absence of spatial dependence when $\alpha_{p,0} \to -\infty$ \cite{murakami2020memory}. Therefore, $\omega_p(s)$ describes a positively dependent map pattern, which is typically assumed in SVC.

The temporal process $w_p(t_q)$ is also defined based on Moran's I statistic on the $q$-th time axis. It measures the temporal dependence of a vector $\bm{\omega}_{p,q}$ whose elements are $w_p(t_q)|t_q \in \{1,\ldots,T_q\}$ as follows: $I[\bm{\omega}_{p,q}]=c_q\frac{\bm{\omega}_{p,q}\bm{M}_q\bm{K}_q\bm{M}_q\bm{\omega}'_{p,q}}{\bm{\omega}_{p,q}\bm{M}_q\bm{\omega}'_{p,q}}$. $c_q$ is a scaling factor, $\bm{K}_q$ is a temporal kernel matrix with zero diagonals, and $\bm{M}_q=\bm{I}_q -\bm{1}_q\bm{1}_q/T_q$, where $\bm{I}_q$ and is an identity matrix and $\bm{1}_q$ is a vector of ones. As with the spatial eigenvectors, the temporal eigenpairs are extracted as $\bm{M}_q\bm{K}_q\bm{M}_q \to \bm{E}_q \bm{\Lambda}_q \bm{E}'_q$, where $\bm{\Lambda}_q$ is a diagonal matrix whose elements are the positive eigenvalues $\lambda_{l_q},\ldots,\lambda_{L_q}$ and 
$\bm{E}_q$ is a matrix of the $L_q$ corresponding eigenvectors, which describe positively dependent temporal patterns measured by the Moran's I statistic.

The temporal process is specified as follows:
\begin{equation} 
w_p(t_q)= \bm{e}(t_q) \bm{\gamma}_{p,q}, \hspace{1cm} \bm{\gamma}_{p,q} \sim N(\bm{0},\tau_{p,q}^2 \bm{\Lambda}_q^{\alpha_{p,q}} ),\label{eq_w_t}
\end{equation}
where $\bm{e}(t_q)$ is the $t_q$-th row of the $\bm{E}_q$ matrix. This process depends on the variance parameter $\tau_{p,q}^2$ and the smoothness parameter $\alpha_{p,q}$, which controls the intensity of temporal dependence measured by $I[\omega_p(t_q)]$.

The spatio-temporal process is specified as follows:
\begin{equation}
w_p(s,t_q)= \left(\bm{e}(s) \otimes \bm{e}(t_q) \right) \bm{\gamma}_{p,q(st)}, \hspace{0.2cm}
\bm{\gamma}_{p,q(st)} \sim N\left(\bm{0}, \tau_{p,q(st)}^2(\bm{\Lambda} \otimes \bm{\Lambda}_q)^{\alpha_{p,q(st)}} \right),\label{eq_w_int}
\end{equation} 
where $\otimes$ represents the Kronecker product. The pattern depends on the variance parameter $\tau_{p,q(st)}^2$ and another parameter $\alpha_{p,q(st)}$ measuring the strength of spatio-temporal dependence measured by the following Moran's I statistic: 
\begin{equation}
I[\bm{\omega}_{p,q(st)}]=c_{q(st)}\frac{\bm{\omega}_{p,q(st)}\bm{M}_{q(st)}\bm{K}_{q(st)}\bm{M}_{q(st)}\bm{\omega}'_{p,q(st)}}{\bm{\omega}_{p,q(st)}\bm{M}_{q(st)}\bm{\omega}'_{p,q(st)}}.\label{eq_moran_st}
\end{equation}
    $\bm{\omega}_{p,q(st)}$ is a vector whose elements are $w_p(s,t_q)|s\in \{1,\ldots,S\},t_q \in \{1,\ldots,T_q\}$, $\bm{K}_{q(st)}=\bm{K} \otimes \bm{K}_q$ is a space-time kernel matrix and $\bm{M}_{q(st)}=\bm{M} \otimes \bm{M}_q$.\footnote{$\bm{M}_{q(st)}\bm{K}_{q(st)}\bm{M}_{q(st)}=(\bm{M} \otimes \bm{M}_q)(\bm{K} \otimes \bm{K}_q)(\bm{M} \otimes \bm{M}_q)=(\bm{MKM}) \otimes (\bm{M}_q \bm{K}_q \bm{M}_q )$. By applying eigen-decomposition to $\bm{MKM}$ and $\bm{M}_q \bm{K}_q \bm{M}_q $, respectively, $\bm{M}_{q(st)}\bm{K}_{q(st)}\bm{M}_{q(st)}$ is approximated by $(\bm{E\Lambda E'}) \otimes (\bm{E_q\Lambda_q E_q'})=(\bm{E} \otimes \bm{E}_q)(\bm{\Lambda} \otimes \bm{\Lambda}_q)(\bm{E} \otimes \bm{E}_q)'$ . The columns of the $\bm{E} \otimes \bm{E}_q$ matrix are mutually orthogonal in the case of longitudinal data, which we assumed, and $\bm{\Lambda} \otimes \bm{\Lambda}_q$ is a diagonal matrix. Therefore, the diagonal elements $\lambda_l \lambda_{l_q}>0$, where $l\in \{1,,\ldots,L\},l_q\in \{1,,\ldots,L_q\}$, represent the positive eigenvalues, and the columns of the $\bm{E} \otimes \bm{E}_q$ matrix represent the corresponding eigenvectors of the $\bm{M}_{q(st)}\bm{K}_{q(st)}\bm{M}_{q(st)}$ matrix. Therefore, similar to the spatial case, these eigen-pairs, which are used in Eq.(\ref{eq_w_int}), are available to describe the process explained by spatio-temporal Moran's I statistic (Eq. \ref{eq_moran_st}).} 

Following \cite{murakami2020scalable}, the $(s,s')$-th element of the spatial kernel matrix $\bm{K}$ is given by $k_{s,s'}=exp(-d_{s,s'}/r_s)$ to model exponentially decaying spatial dependence, where $r_s$ is the longest edge length of the minimum spanning tree (MST) connecting the sample sites (see \cite{dray2006spatial}). The $(t_q,t_q')$-th element of $\bm{K}_q$ is given by $k_{t_q,t_q'}=exp(-d_{t_q,t_q'}/r_q)$, where $r_q$ is the largest time difference separating the sampled time points, which equals the longest edge length of the MST connecting the sampled time points. For the cyclic temporal process (e.g., weekly cycle), the difference $d_{t_q,t_q'}$ is evaluated on the cyclic axis. The elements of $\bm{K}_{q(st)}$ are given by a product kernel $k_{(s,t_q),(s',t_q')}=exp(-d_{s,s'}/r_s)exp(-d_{t_q,t_q'}/r_q)$, which is widely used for spatio-temporal modeling (see Section \ref{sec_intro}).
 
\subsection{Property of the model}
Although both our STVC model and GAM consider basis functions for modeling STVCs, the former has several advantages as follows. First, the number of basis functions/eigenvectors is automatically determined by the number of positive eigenvalues, explaining positive dependence. Second, the proposed model explicitly estimates the strength of spatial and/or temporal dependence measured by Moran's I statistics based on the $\alpha$ parameters, while GAMs implicitly assume $\alpha=0$. Thus, our specification is more appropriate for modeling STVCs. Notably, the eigen-decomposition can be approximated for large samples, as described in \cite{murakami2019eigenvector}.

Our STVC model (Eq. \ref{eq_data}) is rewritten by substituting Eqs.(\ref{eq_beta} - \ref{eq_w_int}) as follows:
\begin{equation}
\begin{split}
y(s,\bm{t})=&\sum_{p=1}^P x_p(s,\bm{t})b_p +\underbrace{\sum_{p=1}^P x_p(s,\bm{t}) \Bigl[\bm{e}(s),\bm{e}(t_1),\ldots,\bm{e}(t_Q) \Bigr]
\bm{\gamma}_0}_{\text{(i) and (ii)}} + \\
& \underbrace{\sum_{p=1}^P x_p(s,\bm{t}) \Bigl[ \bm{e}(s) \otimes \bm{e}(t_1),\ldots,  \bm{e}(s) \otimes \bm{e}(t_Q) \Bigr]
\begin{bmatrix}
    \bm{\gamma}_{p,1(st)} \\
    \vdots \\
    \bm{\gamma}_{p,q(st)}
\end{bmatrix}}_{\text{(iii)}}+ \epsilon(s,\bm{t}),   \label{eq_all}
\end{split}    
\end{equation}
where $\bm{\gamma}_0=[\bm{\gamma}'_{p,0}, \bm{\gamma}'_{p,1},
    \ldots, \bm{\gamma}'_{p,Q}]'$. The first two terms explain main/non-interaction effects (i) and (ii), and the third term explains (iii) space-time interactions.

Equation (\ref{eq_all}) consists of the following basis functions:
\begin{align}
    \text{Main (i) - (ii)} \hspace{0.3cm}:&\hspace{0.1cm}x_p(s,\bm{t})\bm{e}(s),\hspace{0.2cm}x_p(s,\bm{t})\bm{e}(t_1),\hspace{0.2cm}\ldots,\hspace{0.2cm}x_p(s,\bm{t})\bm{e}(t_Q)  \label{eq_bmain}\\  
    \text{Interaction (iii)}:&\hspace{0.1cm}x_p(s,\bm{t})\bigl(\bm{e}(s) \otimes \bm{e}(t_1)\bigr),\hspace{0.2cm}\ldots,\hspace{0.2cm}x_p(s,\bm{t})\bigl(\bm{e}(s) \otimes \bm{e}(t_Q)\bigr)\label{eq_bint}
\end{align}
The main terms (Eq. \ref{eq_bmain}) contain $P(L+\sum_{q=1}^Q L_q)$ basis functions, and the interaction terms (Eq. \ref{eq_bint}) contain $PL\sum_{q=1}^Q L_q$ basis functions. The latter can be quite large. For example, in Section \ref{sec_app1}, we apply our model to a crime analysis assuming $P=5, M=2, L=35, L_1=47$, and $L_2=23$. The resulting numbers of basis functions are 530 for the main and 12,250 for the interaction terms. The latter is considerably larger than typical settings of GAMs, assuming at most thousands of basis functions. In addition, the number of the variance parameters (i.e., $\tau_{p,q}^2, \tau_{p,q(st)}^2$, $\alpha_{p,q}, \alpha_{p,q(st)}$) in our model equals $2P(1+2Q)$, which is considerably larger than conventional GAMs and STVC models. For instance, GTWR considers only two hyper-parameters representing spatial and temporal bandwidths. The estimation can be computationally intractable.

To lighten the computational cost, we select $K_0$ main terms from Eq. (\ref{eq_bmain}) and $K$ interaction terms from Eq. (\ref{eq_bint}). The model with the selected terms is expressed as follows:
\begin{equation}
    \bm{y}=\bm{Xb}+\bm{Z_K}\bm{\gamma_K}+\bm{\epsilon}, \hspace{0.5cm}\bm{\gamma_K} \sim N(\bm{0}, \bm{V}(\bm{\Theta_K})), \hspace{0.5cm} \bm{\epsilon} \sim N(\bm{0},\sigma^2\bm{I}) \label{eq_mat}
\end{equation}
where $\bm{y}$  ($N \times 1$) is a vector of explained variables, $\bm{X}$ ($N \times P$) is a matrix of explanatory variables,
\begin{equation}
\bm{b}=\begin{bmatrix} b_1 \\ \vdots \\ b_P \end{bmatrix}, \hspace{0.1cm}
\bm{\gamma_K}=
\begin{bmatrix} \bm{\gamma}_0 \\ \bm{\gamma}_1 \\ \vdots \\ \bm{\gamma}_K \end{bmatrix}, \hspace{0.1cm}
\bm{V}(\bm{\Theta_K})=
\begin{bmatrix}
\bm{V}(\bm{\Theta}_0) & \bm{O} & \cdots & \bm{O} \\
\bm{O} & \bm{V}(\bm{\theta}_1) & \cdots & \bm{O} \\
\vdots & \vdots & \ddots & \vdots \\
\bm{O} & \bm{O} & \cdots & \bm{V}(\bm{\theta}_K) \\
\end{bmatrix}.
\end{equation}
$\bm{O}$ represents matrix of zeros, and $\bm{\Theta_K}=\{\bm{\Theta}_{0},\bm{\theta}_{1},\ldots,\bm{\theta}_{K}\}$. $\bm{Z_K}=[\bm{Z}_0, \bm{Z}_1,\ldots,\bm{Z}_K]$ with a size of $N \times \tilde{L}$, where $\tilde{L}$ is the total number of basis functions/eigenvectors. $\bm{Z}_0$ is a matrix of the basis functions explaining the $K_0$ main effects selected from Eq. (\ref{eq_bmain}), $\bm{\Theta}_{0}$ is the set of the corresponding variance parameters, and $\bm{V}(\bm{\Theta}_0)$ is a block diagonal matrix whose elements are the variance matrices $\tau_{p,q}^2 \Lambda_{p,q}^{p,q}$ of the $K_0$ selected effects. For example, if $x_3(s,\bm{t})\bm{e}(s)$ and $x_1(s,\bm{t})\bm{e}(t_2)$ are the selected main effects, each row of $\bm{Z}_0$ consists of $[x_3(s,\bm{t})\bm{e}(s), x_1(s,\bm{t})\bm{e}(t_2)]$,  $\bm{\Theta}_0=\{\tau_{3,0}^2,\alpha_{3,0},\tau_{1,2}^2,\alpha_{1,2}\}$, and $\bm{V}(\bm{\Theta}_0)=\begin{bmatrix}
   \tau_{3,0}^2 \bm{\Lambda}^{\alpha_{3,0}} & \bm{O} \\
    \bm{O} & \tau_{1,2}^2 \bm{\Lambda}_2^{\alpha_{1,2} }
\end{bmatrix}$. $\bm{Z}_k$ is a matrix of the basis functions describing the $k$-th selected interaction effect from Eq. (\ref{eq_bint}), and $\bm{\theta}_k$ is the corresponding variance parameters. For instance, if $x_4(s,\bm{t})\bigl(\bm{e}(s) \otimes \bm{e}(t_3)\bigr)$ is the $k$-th selected effect, each row of $\bm{Z}_k$ is given by $x_4(s,\bm{t})\bigl(\bm{e}(s) \otimes \bm{e}(t_3)\bigr)$, $\bm{\theta}_k=\{ \tau^2_{4,3(st)}, \alpha_{4,3(st)} \}$, and $\bm{V}(\bm{\theta}_k)=\tau^2_{4,3(st)}(\bm{\Lambda}\otimes \bm{\Lambda}_3)^{\alpha_{4,3(st)}}$.

\subsection{Estimation}\label{subsec_select}
To estimate the STVCs while addressing the computational burden and over-fitting associated with the large basis dimension, we combine a fast likelihood maximization method \cite{murakami2020scalable} with a model selection inspired by reluctant interaction modeling \cite{yu2019reluctant}. Following the latter, we distinguish between the main/non-interaction and interaction terms, which can slow down computation. 

\subsubsection{Marginal likelihood}\label{subsec_loglik}
Eq. (\ref{eq_mat}) is identical to the conventional linear mixed model. As derived in Appendix \ref{app_lik}, the following marginal log-likelihood\footnote{It is derived by marginalizing $\bm{b}$ and $\bm{\gamma_K}$ from the full likelihood. As explained in Section 6.2.6 of \cite{wood2017book}, it is a marginal log-likelihood in a Bayesian fashion, which is obtained by considering $\bm{\gamma_K} \sim N(\bm{0},\bm{V(\Theta_K)})$ as a Gaussian prior. It is a type of restricted likelihood.} with respect to the variance parameters $\bm{\Theta_K}=\{\bm{\Theta}_{0},\bm{\theta}_{1},\ldots,\bm{\theta}_{K}\}$ is readily obtained as follows (see \cite{murakami2019spatially}):
\begin{equation}
l(\bm{\Theta_K})= -\frac{1}{2}\log|\bm{H_{K,K}}|-\frac{N-P}{2} \left(1+\log \bigg(2\pi\frac{ \hat{\bm{\epsilon}}'\hat{\bm{\epsilon}} + \hat{\bm{u}}'\hat{\bm{u}} }{N-P}  \bigg) \right),\label{eq_loglik}
\end{equation}
where $\bm{V}(\bm{\Theta_K})^{1/2} \hat{\bm{u}}_{\bm{K}}=\bm{\hat{\gamma}_K}$, $\hat{\bm{\epsilon}}=\bm{y}-\bm{X\hat{b}}-\bm{Z_K}\bm{V}(\bm{\Theta_K})^{1/2} \hat{\bm{u}}_{\bm{K}}$ and
\begin{equation}
\begin{bmatrix}
  \hat{\bm{b}} \\ \hat{\bm{u}}_{\bm{K}} \\
\end{bmatrix} = \bm{H_{K,K} }^{-1}\bm{h_{K,y}}.\label{eq:penPLS}
\end{equation}
$\bm{H_{K,K}}=\begin{bmatrix}
\bm{X'X} & \bm{G_{K,X}}' \\
\bm{G_{K,X}} &\bm{G_{K,K}}+\bm{I}
\end{bmatrix}$ and $\bm{h_{K,y}}=
\begin{bmatrix}
\bf{X}'\bf{y} \\ \bm{G_{K,y}}
\end{bmatrix}$ where
\begin{equation*}
\bm{G_{K,K}}=
\begin{bmatrix}
\bm{G}_{0,0} & \bm{G}_{0,1} & \ldots & \bm{G}_{0,K} \\
\bm{G}_{1,0} & \bm{G}_{1,1} & \ldots & \bm{G}_{1,K} \\
\vdots & \vdots & \ddots & \vdots \\
\bm{G}_{K,0} & \bm{G}_{K,1} & \ldots & \bm{G}_{K,K} \\
\end{bmatrix}, \hspace{1cm} 
\bm{G_{K,X}}=\begin{bmatrix}
\bm{G}_{1,\bm{X}}\\ \bm{G}_{2,\bm{X}} \\ \vdots \\ \bm{G}_{K,\bm{X}}
\end{bmatrix},
\end{equation*}
in which $\bm{G}_{k,\tilde{k}}=\sigma^{-2}\bm{V}(\bm{\theta}_k)^{1/2}\bm{Z}'_{k}\bm{Z}_{\tilde{k}}\bm{V}(\bm{\theta}_{\tilde{k}})^{1/2}$ and $\bm{G}_{k,\bm{X}}=\sigma^{-1}\bm{V}(\bm{\theta}_k)^{1/2}\bm{Z}'_k\bm{X}$. $\bm{G_{K,y}}$ is defined similarly for $\bm{y}$.

For model selection, we will use the marginal Bayesian information criterion (BIC), which is defined as follows (see \cite{delattre2014note},\cite{sakamoto2019bias}):
\begin{equation}
    BIC(\bm{\Theta}_{\bm{K}})=-2l(\bm{\Theta}_{\bm{K}})+\mathrm{log}(N)(P+\mathrm{dim}(\bm{\Theta}_{\bm{K}})+1), \label{eq_bic}
\end{equation}
where $\mathrm{dim}(\bm{\Theta}_{\bm{K}})$ denotes the number of elements in $\bm{\Theta}_{\bm{K}}$. \cite{murakami2020scalable} demonstrated the accuracy of an SVC model selection based on the marginal BIC.

Once the parameters in $\bm{\Theta}_{\bm{K}}$ are estimated, the coefficient estimates $\hat{\bm{b}}$ and $\bm{\hat{\gamma}_K}$ are readily obtained using Eq. (\ref{eq:penPLS}). These estimates are then used to obtain the predictive mean of the STVCs. For example, if the spatial pattern, the 2nd and 3rd temporal patterns, and the 1st and 2nd interaction patterns are selected for the $p$-th STVC, the predictive mean yields
\begin{equation}
    \hat{\beta_p}(s,\bm{t})=\hat{b}_p + \bm{e}(s)\hat{\bm{\gamma}}_p + \sum_{q=2}^3 \bm{e}(t_q)\hat{\bm{\gamma}}_{p,q} + \sum_{q=1}^2 (\bm{e}(s) \otimes \bm{e}(t_q))\hat{\bm{\gamma}}_{p,q(st)},\label{eq_svc_estimate}
\end{equation}
where $\hat{b}_p$ is the $p$-th element of $\hat{\bm{b}}$ and $\hat{\bm{\gamma}}_p$, $\hat{\bm{\gamma}}_{p,q}$ $\hat{\bm{\gamma}}_{p,q(st)}$ are elements of $\bm{\hat{\gamma}_K}$.

\subsubsection{Estimation and model selection procedure}
The log-likelihood contains  $\bm{H_{K,K}}^{-1}$, whose size equals $(P+\tilde{L})\times (P+\tilde{L})$, and $|\bm{H_{K,K}}|$. Their computational complexities rapidly grow in an order of $(P+\tilde{L})^3$ . Because of the large basis dimension in the interaction terms, their calculations can be very slow. 

To avoid explicit calculations of these two, we propose to select the main effects first, and then select the interaction terms while updating $\bm{H_{K,K}}^{-1}$ and $|\bm{H_{K,K}}|$ in each step. The calculation procedure is summarized as follows:
\begin{enumerate}
\item Estimate and select the main effects only model $M_0$: $\bm{y}=\bm{Xb}+\bm{Z}_0\bm{\gamma}_0+\bm{\epsilon}$, $\bm{\gamma}_0 \sim N(\bm{0},\bm{V}(\bm{\Theta}_0))$, $\bm{\epsilon} \sim N(\mathbf{0}, \sigma^2 \bm{I})$ following the process described in  \cite{murakami2020scalable}. The calculation procedure is summarized as follows:
\begin{enumerate}
    \item Initialize $\bm{\hat{\Theta}}_0$ and evaluate $BIC(\hat{\bm{\Theta}}_0)$. 
    \item The following steps are repeated for each $p \in \{0,1,\ldots,P\}$ and $q \in \{0,1,\ldots,Q\}$ until the marginal BIC does not improve:
\begin{enumerate}
    \item Partition the parameters as $\bm{\hat{\Theta}}_0=\{ \bm{\hat{\theta}}_{0(p,q)},\bm{\hat{\Theta}}_{0(-p,q)}\}$, where $\bm{\hat{\theta}}_{0(p,q)}=\{\hat{\alpha}_{p,q},\hat{\tau}^2_{p,q}\}$. Estimate $\hat{\bm{\theta}}^{cand}_{0(p,q)}$ by maximizing the log-likelihood $l(\bm{\theta}_{0(p,q)}, \bm{\hat{\Theta}}_{0(-p,q)} )$ given $\bm{\hat{\Theta}}_{0(-p,q)}$. Subsequently, evaluate $BIC(\hat{\bm{\theta}}^{cand}_{0(p,q)}, \bm{\hat{\Theta}}_{0(-p,q)})$ using Eq.(\ref{eq_bic}). 
    \item Evaluate $BIC(\bm{\hat{\Theta}}_{0(-p,q)} )$ of the model that eliminates the $(p,q)$-th main effect from the model $M_0$. 
    \item If $BIC(\hat{\bm{\theta}}^{cand}_{0(p,q)}, \bm{\hat{\Theta}}_{0(-p,q)})<BIC(\bm{\hat{\Theta}}_{0(-p,q)})$, $\bm{\hat{\Theta}}_0=\{\hat{\bm{\theta}}^{cand}_{0(p,q)}, \bm{\hat{\Theta}}_{0(-p,q)}\}$ is the updated parameter estimates. Otherwise, the model $M_0$ is updated by assuming $\bm{\hat{\Theta}}_0=\bm{\hat{\Theta}}_{0(-p,q)}$ and removing the basis functions explaining the ($p,q$)-th main effect from $\bm{Z}_0$.    
\end{enumerate}
\end{enumerate}
\item Initialize $\kappa^{cand}=PM$ and $K=0$, as well as $M_{\bm{K}}=M_0$, $\hat{\bm{\Theta}}_{\bm{K}}
=\hat{\bm{\Theta}}_0$, and $\bm{Z_K}=\bm{Z}_0$. Calculate $\bm{H_{K,K}} =\bm{H}_{0,0}$ using $\bm{Z}_0$, and evaluate $\bm{H_{K,K}}^{-1}$ and $|\bm{H_{K,K}}|$. In the subsequent step, use them to evaluate the likelihood.
\item Iterate the following procedure for each $\kappa \in \{ 1,\ldots, \kappa^{cand}\}$, and select $\hat{\kappa}$ minimizing the marginal BIC value:
\begin{enumerate}
    \item Estimate $\bm{\theta}_\kappa$ by maximizing the likelihood of the following model $M_{\kappa}$ composed of the already selected terms and the $\kappa$-th interaction term:
    \begin{equation}
    \begin{split}
    M_{\kappa}:\bm{y}&=\bm{Xb}+\bm{Z_K}\bm{\gamma}_{\bm{K}}+\bm{Z}_\kappa \bm{\gamma}_{\kappa}+\bm{\epsilon},\hspace{0.5cm} \bm{\epsilon} \sim N(\mathbf{0},\sigma^2 \bm{I}), \\
    \bm{\gamma}_{\bm{K}} &\sim N(\bm{0}, \bm{V}(\bm{\hat{\Theta}_K})), \hspace{1cm} \bm{\gamma}_{\kappa} \sim N(\bm{0}, \bm{V}(\bm{\theta}_\kappa)).
    \end{split}
    \end{equation} 
    In this step, $\hat{\bm{\Theta}}_{\bm{K}}$ is fixed, and $\bm{\theta}_\kappa$ is estimated by numerically maximizing the log-likelihood.
    \item Evaluate the marginal BIC of the model $M_\kappa$.
\end{enumerate}
\item If the BIC value of the selected model $M_{\hat{\kappa}}$ is worse than $M_{\bm{K}}$, $M_{\bm{K}}$ is the final model. Otherwise, $M_{\hat{\kappa}}$ is the updated model. In the latter case, calculate the following quantities required to evaluate the updated log-likelihood value: $M_{\bm{K}+1} = M_{\hat{\kappa}}$, $\bm{\hat{\Theta}}_{\bm{K}+1}=\{\bm{\hat{\Theta}}_{\bm{K}},\hat{\bm{\theta}}_{\hat{\kappa}}\}$,  $\bm{Z}_{\bm{K}+1}=[\bm{Z_K},\bm{Z}_{\hat{\kappa}}]$, and 
\begin{equation}
\bm{H}_{\bm{K}+1,\bm{K}+1}=
\begin{bmatrix}
    \bm{H_{K,K}} & \bm{G}_{\bm{K},\hat{\kappa} } \\
    \bm{G}'_{\bm{K},\hat{\kappa} }& \bm{G}_{\hat{\kappa},\hat{\kappa} }
\end{bmatrix}. \label{eq_hmat}
\end{equation}
where $\bm{G}_{\bm{K},\hat{\kappa} } = [\bm{G}'_{1,\hat{\kappa} },\ldots, \bm{G}'_{K,\hat{\kappa} }]'$ and $\bm{G}_{k,\hat{\kappa} }$ is similarly defined as $\bm{G}_{k,k'}$.
Subsequently, evaluate $\bm{H}_{\bm{K}+1,\bm{K}+1}^{-1}$ and $|\bm{H}_{\bm{K}+1,\bm{K}+1}|$ following the fast procedure explained in Section \ref{sec_inv_det}. 
\item Exclude $\hat{\kappa}$ from the candidate set as $\kappa \in \{1, \ldots, \kappa^{cand} \}\backslash \hat{\kappa}$.
\item $K \to K + 1$
\item Iterate steps 3 - 5 until BIC converges. $M_{\bm{K}}$ is the final model.
\end{enumerate}
We used the marginal BIC (Eq. \ref{eq_bic}) following \cite{murakami2020scalable}, suggesting the accuracy of the BIC-based model selection in he case of an SVC model.

\subsubsection{Inverse and determinant calculation}\label{sec_inv_det}
The estimation in Step 3 (a) iteratively evaluates $\bm{H}_{\bm{K}+1,\bm{K}+1}^{-1}$ to find the maximum likelihood estimates, which is computationally demanding. The inversion can be expressed by applying Schur's lemma to Eq. (\ref{eq_hmat}), as follows: 
\begin{equation}
    \bm{H}_{\bm{K}+1,\bm{K}+1}^{-1}=
    \begin{bmatrix}
    \bm{H_{K,K}}^{-1} + \bm{H_{K,K}}^{-1}
    \bm{G}_{\bm{K},\kappa}\bm{Q}^{-1}_{\kappa,\kappa}\bm{G}_{\bm{K},\kappa}'\bm{H_{K,K}}^{-1}& 
    -\bm{H_{K,K}}^{-1}
    \bm{G}_{\bm{K},\kappa}\bm{Q}^{-1}_{\kappa,\kappa}\\
    -\bm{Q}^{-1}_{\kappa,\kappa}\bm{G}_{\bm{K},\kappa}'\bm{H_{K,K}}^{-1} &
    \bm{Q}^{-1}_{\kappa,\kappa}

    \end{bmatrix}\label{eq_solve}
\end{equation}
where $\bm{Q}_{\kappa,\kappa}= \bm{G}_{\kappa,\kappa}-\bm{G}_{\bm{K},\kappa}'\bm{H_{K,K}}^{-1}\bm{G}_{\bm{K},\kappa}$ with a size of $L_{\kappa}\times L_{\kappa}$, where $L_{\kappa}$ is the number of basis functions in the $\kappa$-th interaction term. Since $\bm{H_{K,K}}^{-1}$, the heaviest part in Eq. (\ref{eq_solve}), has already been evaluated in the previous iteration, $\bm{Q}_{\kappa,\kappa}$ is the only matrix to be iteratively inverted in Step 3(a). The computational complexity equals $O(L_{\kappa}^3)$, which is considerably smaller than the complexity $O((P+\tilde{L})^3)$ in the direct calculation of $\bm{H}_{\bm{K}+1,\bm{K}+1}^{-1}$ .

The determinant $|\bm{H}_{\bm{K}+1,\bm{K}+1}|$ also requires a heavy computation. To avoid the direct evaluation, we use the following equality, which is obtained by applying another Schur's formula to Eq. (\ref{eq_hmat}):
\begin{equation}
|\bm{H}_{\bm{K}+1,\bm{K}+1}|=|\bm{H_{K,K}}||\bm{Q}_{\kappa,\kappa}|.
\end{equation}
Since $|\bm{H_{K,K}}|$ was evaluated in the previous iteration and $\bm{Q}_{\kappa,\kappa}$ was evaluated above, the additional computational cost is quite small.

In summary, we proposed an STVC modeling method that sequentially adds interaction terms. Because a small matrix $\bm{Q}_{\kappa,\kappa}$ is inverted in each sequence instead of the large matrix $\bm{H_{K,K}}$, the model estimation is computationally efficient. 

\section{Monte Carlo experiments}\label{sec_mc}

This section demonstrates three Monte Carlo experiments. Section \ref{subsec_basic} assesses whether the proposed method successfully estimates the true patterns while avoiding over-fitting. Section \ref{subsec_compare_gam} compares the proposed method with GAM, which is an alternative capable of modeling all the patterns considered in our model. Section \ref{subsec_compare_gwr} compares the proposed method with GTWR. Following GTWR, this section considers non-cyclic patterns only. 

\subsection{Basic comparison}\label{subsec_basic}
\subsubsection{Outline}
This section considers the following regression model:
\begin{equation}
\begin{split}\label{sim_basic}
y(s,\bm{t})&=\beta_1(s,\bm{t})+ x_2(s,\bm{t})\beta_2(s,\bm{t})+x_3(s,\bm{t})\beta_3(s,\bm{t})+\epsilon(s,\bm{t})\hspace{0.5cm}\epsilon(s,\bm{t})\sim N(0,1),
\end{split}
\end{equation}
where $x_p(i,\bm{t})\sim N(0,1)$ and $\bm{t}=\{t_1,t_2\}$. The following specifications are considered for the three coefficients: 
\begin{equation}
\begin{split}
\mathrm{LM} \quad \beta_p(s,\bm{t})&=b_p, \\
\mathrm{S} \quad \beta_p(s,\bm{t})&=b_p+v_p[w_p(s)], \\
\mathrm{ST} \quad \beta_p(s,\bm{t})&=b_p+v_p[w_p(s)+w_p(t_1)], \\
\mathrm{ST_{int}} \quad \beta_p(s,\bm{t})&=b_p+v_p[w_p(s)+w_p(t_1)+w_p(s,t_1)],\\
\mathrm{STc} \quad \beta_p(s,\bm{t})&=b_p+v_p[w_p(s)+ w_p(t_1) +w_p(t_2)], \\
\mathrm{STc_{int}} \quad \beta_p(s,\bm{t})&=b_p+v_p[w_p(s)+w_p(t_1)+w_p(t_2) +w_p(s,t_1) +w_p(s,t_2)], \label{eq_stc_int}
\end{split}
\end{equation}
where $\{b_1,b_2,b_3\}=\{1,2,-0.5\}$  and  $[\cdot]$ denotes standardization to zero mean and variance one. $w_p(t_1)$ and $w_p(t_2)$ are non-cyclic and cyclic temporal processes, respectively, and $w_p(s,t_1)$ and $w_p(s,t_2)$ are their corresponding interaction terms. $w_p(s), w_p(t_q), w_p(s,t_q)$ are randomly sampled using Eqs. (\ref{eq_w_s} - \ref{eq_w_int}), assuming  $\alpha_p=\alpha_{p,q}=\alpha_{p,q(st)}=1$  and  $\tau_p^2=\tau_{p,q}^2=\tau_{p,q(st)}^2$  for simplicity \footnote{The absolute values of $\tau_p^2,\tau_{p,q}^2,\tau_{p,q(st)}^2$ do not change the result because the variances are rescaled by the $[\cdot]$ operator}.

S and T in Eq. (\ref{eq_stc_int}) represent space and time, respectively, $c$ represents circular time, and the subscript $int$ denotes space-time interaction. The standard deviations of the three coefficients are given by $(v_1, v_2, v_3)=(1,2,1)$. We refer to $\beta_2(s,\bm{t})$, which explains a larger amount of variation, the strong coefficient, and $\beta_3(s,\bm{t})$, which explains a smaller variation, the weak coefficient. 

Synthetic samples are generated from Eq. (\ref{sim_basic}) at 200 locations over 40 regular time points, assuming a circular period of 10. The spatial coordinates are generated from two uniform distributions. The samples are generated under the following scenarios for the true coefficients $\beta_1(s,\bm{t}),\beta_2(s,\bm{t}),\beta_3(s,\bm{t})$: (I) $\rm{ST, STc_{int}, LM}$, (II) $\rm{ST, LM, STc_{int}}$, (III) $\rm{STc_{int}, LM, LM}$.  

In each scenario, we fit six models assuming $\rm{LM, S, ST, ST_{int}, STc, STc_{int}}$ (Eq. \ref{eq_stc_int}) as the full specification for $\beta_p(s,\bm{t})$, while $b_p$ as the minimal specification. For each model, the elements in each coefficient are estimated/selected using our proposed procedure. The model estimation is repeated 400 times for each scenario, and the gap between the true coefficient values and their predictive means (see Eq. \ref{eq_svc_estimate}) is evaluated by the root mean squared error (RMSE).

Among the models, $\rm{STc_{int}}$ considers all discussed patterns. The main objective of this section is to assess whether this model accurately estimates both simple and complex patterns while avoiding over- and under-fitting.

\subsubsection{Result}
\begin{figure}[hbtp]
 \centering
 \includegraphics[scale=0.53]{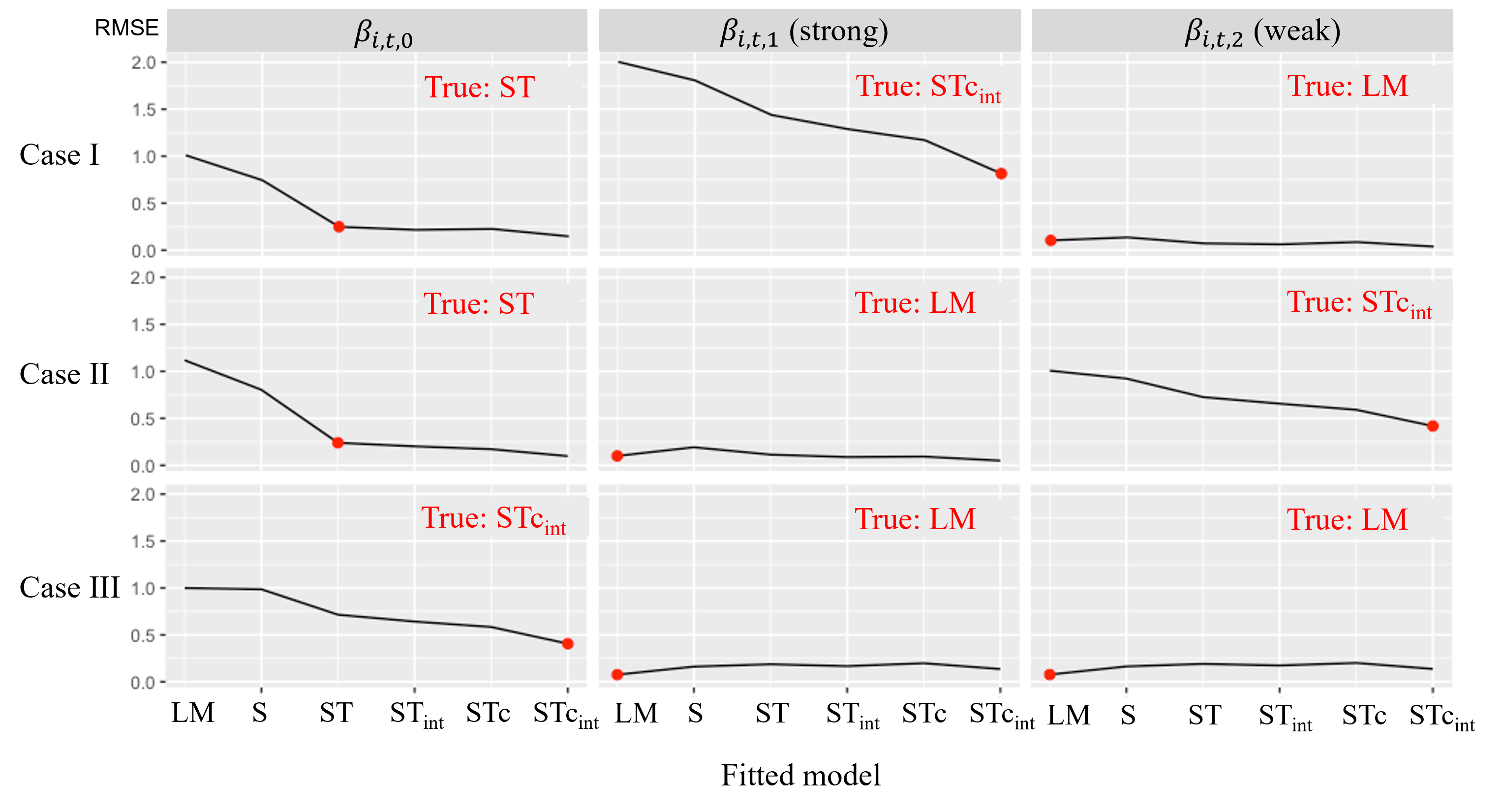}
 \caption{RMSEs of $\beta_1(s,\bm{t})$, $\beta_2(s,\bm{t})$, and $\beta_3(s,\bm{t})$ under scenarios (I - III) for the fitted models. The red dots represent the RMSE values obtained by fitting the true model.} 
 \label{fig_recoveries_hetero}
\end{figure}

Figure\ref{fig_recoveries_hetero} summarizes the RMSEs. In many cases, the RMSE values are the smallest when the fitted and true processes coincide. Interestingly, the full model $\rm{STc_{int}}$ also performs the best in most cases. This result suggests that the proposed method accurately describes the space-time patterns in the regression coefficients, regardless of whether the true pattern is simple or complex.

Table \ref{tab_prop} summarizes the proportions of each element ($w_p(s), w_p(t_1), w_p(t_2), w_p(s,t_1), w_p(s,t_2)$) selected for each coefficient. In all the cases, truly exiting elements are appropriately selected by more than 89 $\%$ of probability. Regarding the constant coefficients, which are $\beta_3(s,\bm{t})$ in case 2 and $\beta_2(s,\bm{t})$ in case 1, the space-time interactions are appropriately estimated as absent in more than 96 $\%$ of the simulations. Thus, the accuracy of our interaction selection method is verified. In case 3, the constant coefficients $\beta_2(s,\bm{t})$ and $\beta_3(s,\bm{t})$ are mistakenly estimated as varying with a certain probability, probably because of the difficulty in distinguishing spatio-temporal variations in the intercept and regression coefficients. Nevertheless, as listed in Table \ref{tab_sd}, the standard deviations of the estimated $\beta_2(s,\bm{t})$ and $\beta_3(s,\bm{t})$ in case 3 are reasonably small. Our STVC model is confirmed to accurately estimate the coefficients with the help of the interaction selection method. Improving our model selection accuracy in case 3  remains an important issue that can be resolved in different ways, for example, by applying Lasso, penalized complexity \cite{simpson2017penalising}, or other penalty.

\begin{table}[hbtp]
\caption{Proportions of each element $w_p(s), w_p(t_1), w_p(t_2), w_p(s,t_1), w_p(s,t_2)$ selected over the Monte Carlo iterations. Existing elements, which should be selected, are shown in bold.}\label{tab_prop}
\begin{center}
\begin{tabular}{cl|rrr|rr} \hline
\multicolumn{1}{l}{} &  & $w_p(s)$ & $w_p(t_1)$ & $w_p(t_2)$ & $w_p(s,t_1)$ & $w_p(s,t_2)$ \\
\hline
\multirow{3}{*}{Case 1} & $\beta_1(s,\bm{t})$ & \textbf{1.00} & \textbf{1.00} & 0.34 & 0.52 & 0.17 \\
 & $\beta_2(s,\bm{t})$ & \textbf{1.00} & \textbf{1.00} & \textbf{1.00} & \textbf{1.00} & \textbf{0.95} \\
 & $\beta_3(s,\bm{t})$ & 0.33 & 0.17 & 0.07 & 0.03 & 0.03 \\
 \hline
\multirow{3}{*}{Case 2} & $\beta_1(s,\bm{t})$ & \textbf{1.00} & \textbf{1.00} & 0.41 & 0.25 & 0.06 \\
 & $\beta_2(s,\bm{t})$ & 0.37 & 0.59 & 0.23 & 0.02 & 0.04 \\
 & $\beta_3(s,\bm{t})$ & \textbf{1.00} & \textbf{1.00} & \textbf{0.99} & \textbf{1.00} & \textbf{0.89} \\
 \hline
\multirow{3}{*}{Case 3} & $\beta_1(s,\bm{t})$ & \textbf{1.00} & \textbf{1.00} & \textbf{1.00} & \textbf{1.00} & \textbf{0.93} \\
 & $\beta_2(s,\bm{t})$ & 0.92 & 0.75 & 0.26 & 0.58 & 0.28 \\
 & $\beta_3(s,\bm{t})$ & 0.94 & 0.81 & 0.38 & 0.58 & 0.28 \\
 \hline
\end{tabular}
\end{center}
\end{table}

\begin{table}[hbtp]
\caption{Mean standard deviation of the true and estimated coefficients. For the latter, the only coefficients that are estimated varying (i.e., not constant) are used for the evaluation.}\label{tab_sd}
\begin{center}
    \begin{tabular}{l|rrr|rrr}
\hline
 & \multicolumn{3}{c|}{True} & \multicolumn{3}{c}{Estimate} \\
 \cline{2-7}
  & $\beta_1(s,\bm{t})$ & $\beta_2(s,\bm{t})$ & $\beta_3(s,\bm{t})$ & $\beta_1(s,\bm{t})$ & $\beta_2(s,\bm{t})$ & $\beta_3(s,\bm{t})$ \\
 \hline
Case 1 & 1.00& 2.00 & 0.00 & 1.01 & 1.81 & 0.09 \\
Case 2 & 1.00& 0.00 & 1.00 & 1.00 & 0.06 & 0.96 \\
Case 3 & 1.00 & 0.00 & 0.00 & 0.90 & 0.14 & 0.14\\ 
\hline
\end{tabular}
\end{center}
\end{table}

\subsection{Comparison with GAM}\label{subsec_compare_gam}
\subsubsection{Outline}
This section compares $\rm{STc_{int}}$ with GAM, a scalable alternative for large samples. We consider two specifications called GAM and GAM2. GAM defines each STVC by [non-cyclic interaction process] + [cyclic interaction process]. These interaction processes are defined by the product of space and temporal basis functions approximating GPs (see \cite{kammann2003geoadditive})\footnote{Since the cyclic temporal basis function is not implemented in the mgcv package, the cyclic basis functions are approximated by acyclic ones.}. GAM2 defines each STVC by [above-mentioned interaction terms] + [pure spatial process] + [non-cyclic temporal process] + [cyclic temporal process]. Thus, GAM2 is more consistent with $\rm{STc_{int}}$. 

GAM and GAM2 are implemented using the R package mgcv (\url{https://cran.r-project.org/web/packages/mgcv/index.html}). Their basis functions are selected using the method of \cite{marra2011practical}, which is implemented in the package. Although we tried to consider the same number of basis functions as our $\rm{STc_{int}}$, the resulting GAMs became too slow to use in the simulation. Therefore, we rely on the default basis dimension settings in mgcv.

The modeling accuracy of these models is compared by fitting them on synthetic data generated from the following STVC model:
\begin{equation}
\begin{split}
y_{i,t}&=\beta_1(s,\bm{t})+ x_2(s,\bm{t})\beta_2(s,\bm{t})+x_3(s,\bm{t})\beta_3(s,\bm{t})+\epsilon(s,\bm{t})\hspace{0.5cm}\epsilon(s,\bm{t})\sim N(0,\sigma^2),\\
\beta_p(s,\bm{t})&=b_p + \tau_p[w_p(s) + w_p(t_1) + w_p(t_2) + w_p(s,t_1) +w_p(t_2)], 
\end{split}
\end{equation}
where the standard deviations of the varying coefficients are $\{\tau_1,\tau_2,\tau_3\}=\{1,2,0.5\}$, meaning that $\beta_2(s,\bm{t})$ is the strong STVC while $\beta_3(s,\bm{t})$ is a weak STVC, similar to the previous experiment. The cyclic time periods for $w_p(t_2)$ and $w_p(s,t_2)$ are set as 20.

100,000 synthetic explained variables were generated at 500 locations at 200 regular time points. The spatial coordinates were generated from two uniform distributions. Randomly selected $N$ samples among them were regarded as observations and the remaining samples were used for predictive accuracy evaluation. The circular period was assumed as 20.

This section considers the small, medium, and large sample scenarios, assuming $N \in \{2000, 10000, 50000\}$, respectively. In each scenario, the models are fit 200 times and their accuracy and computational efficiency are compared. 
 
\subsubsection{Result}\label{subsec_gam_res}

\begin{figure}
    \centering
    \includegraphics[width=12.5cm]{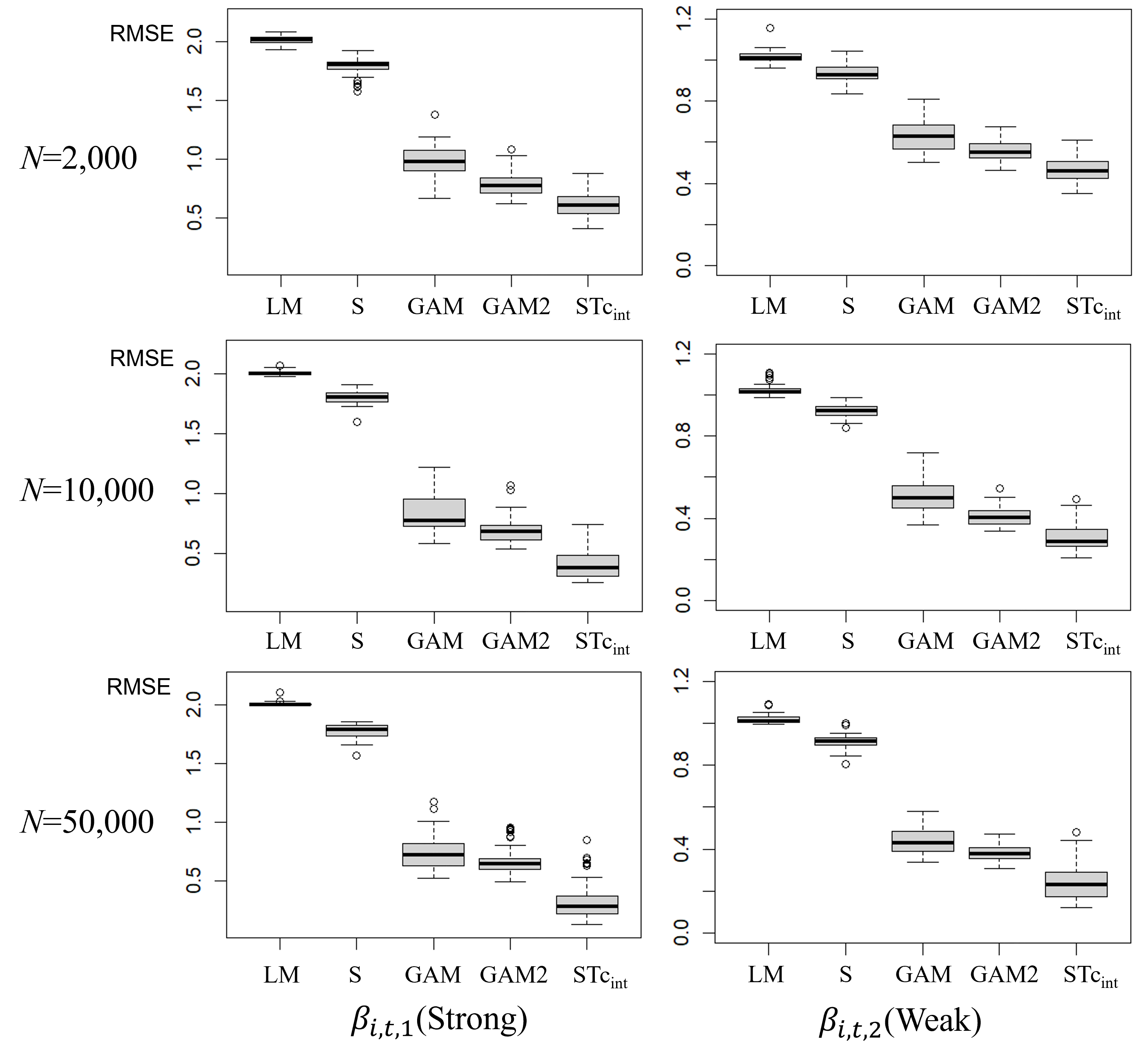}
    \caption{RMSE of the STVC estimates}
    \label{fig_rmse_large}
\end{figure}

\begin{figure}
    \centering
    \includegraphics[width=12cm]{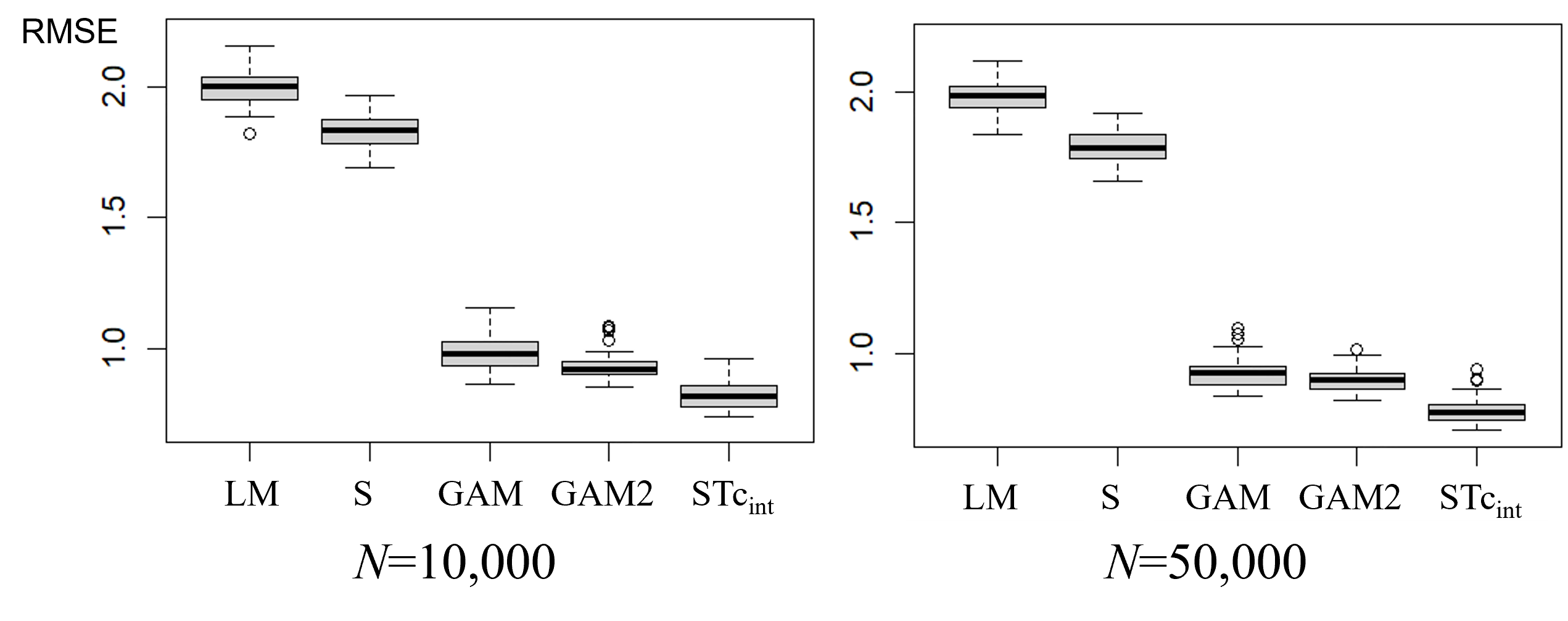}
    \caption{RMSE of the predictive values}
    \label{fig_prmse_large}
\end{figure}

Figure\ref{fig_rmse_large} displays boxplots of the RMSE values for the strong and weak STVCs. For reference, LM and S are also compared. As expected, GAM, GAM2, and $\rm{STc_{int}}$ outperform S owing to their consideration of temporal variations. Among them, RMSEs are smaller for $\rm{STc_{int}}$, GAM2, and GAM, in that order, coinciding with the descending order of their basis dimensions. The accuracy of $\rm{STc_{int}}$ can be attributed to the larger number of basis functions introduced by the proposed procedure. As elaborated later, conventional GAMs struggle with such a large basis dimension due to computational costs. The proposed method efficiently enhances accuracy as the sample size increases from 10000 to 50000, whereas GAMs show less improvement. This underscores the usefulness of our method in improving modeling accuracy for larger samples.
Fig.\ref{fig_prmse_large} compares the out-of-sample predictive accuracy, confirming that our model successfully enhances predictive accuracy.

Table \ref{tab_time} compares the computation time using a 12-core Mac computer with 64 GB of memory. We considered another GAM2 assuming the same number of basis functions as $\rm{STc_{int}}$, denoted as GAM2*. According to the table, $\rm{STc_{int}}$ is significantly faster than GAM2*. The computational efficiency of $\rm{STc_{int}}$ was confirmed. In addition, GAM2* was not available when $N= 2,000$ because of the singular fit. The proposed algorithm is essential for handling high-dimensional STVC models stably.

\begin{table}[hbtp]
  \caption{Average computation time for the first five trials (in seconds). GAM2* could not be estimated when $N$ = 2,000 because of the large basis dimension relative to $N$. Although these models can be parallelized using the mgcv package, the resulting computation times were slower than non-parallel estimation. This is likely due to the large number of basis functions assumed in this study. These computation times are not reported.}\label{tab_time}
  \centering
  \begin{tabular}{r|rrr}
    \hline
    $N$ & S & GAM2* & $\rm{STc_{int}}$ \\
    \hline
    2,000 & 6.8 & NA & 11.4 \\
    10,000 & 8.0 & 2161.5 & 296.6 \\
    50,000 & 19.5 & 2970.2 & 342.9  \\
    \hline
  \end{tabular}
\end{table}


\subsection{Comparison with GTWR}\label{subsec_compare_gwr}
\subsubsection{Outline}
This section compares the proposed method with GWR and GTWR, which are widely used for modeling SVC and STVC, respectively. Following our model, an exponential kernel is used for their local weighting, and the bandwidth is optimized by minimizing the corrected Akaike Information Criterion (AICc). GWR and GTWR were implemented using the R package GWmodel (\url{https://cran.r-project.org/web/packages/GWmodel/index.html}).\footnote{The kernel of GWR is defined by an exponentially-decaying function of spatial distance only. In other words, regardless of the time difference, the local weights on any two locations are the same if they have the same spatial distance from the regression point. The bandwidth is estimated by maximizing the AICc.}

We assume three scenarios for data generation. In alignment with the GTWR assumptions, all scenarios incorporate only non-cyclical time patterns. Scenario I generates synthetic data as follows:
\begin{equation}
\begin{split} \label{eq_sim_gw}
y(s,t)=\beta_1(s,t)&+ x_2(s,t)\beta_2(s,t)+x_3(s,t)\beta_3(s,t)+\epsilon(s,t)\hspace{0.5cm}\epsilon(s,t)\sim N(0,\sigma^2),\\
\beta_p(s,t)=b_p &+ \tau_p[w_p(s) + w_p(t) + w_p(s,t)]. \\
\end{split}
\end{equation}
As in the previous section, $\beta_2(s,t)$ and $\beta_3(s,t)$ are the strong and weak STVCs in Scenario I, respectively. Scenario II replaces the weak STVC with a weak SVC defined as $\beta_3(s,t)=b_2 + \tau_2[w_2(s)]$. Scenario III replaces the strong STVC in Scenario I with a strong SVC. Considering the computational cost of GTWR, this section assumes 1,000 samples observed at 10 regular time points at 100 locations, with coordinates generated by uniform distributions.

\subsubsection{Result}
\begin{figure}
    \centering
    \includegraphics[width=13cm]{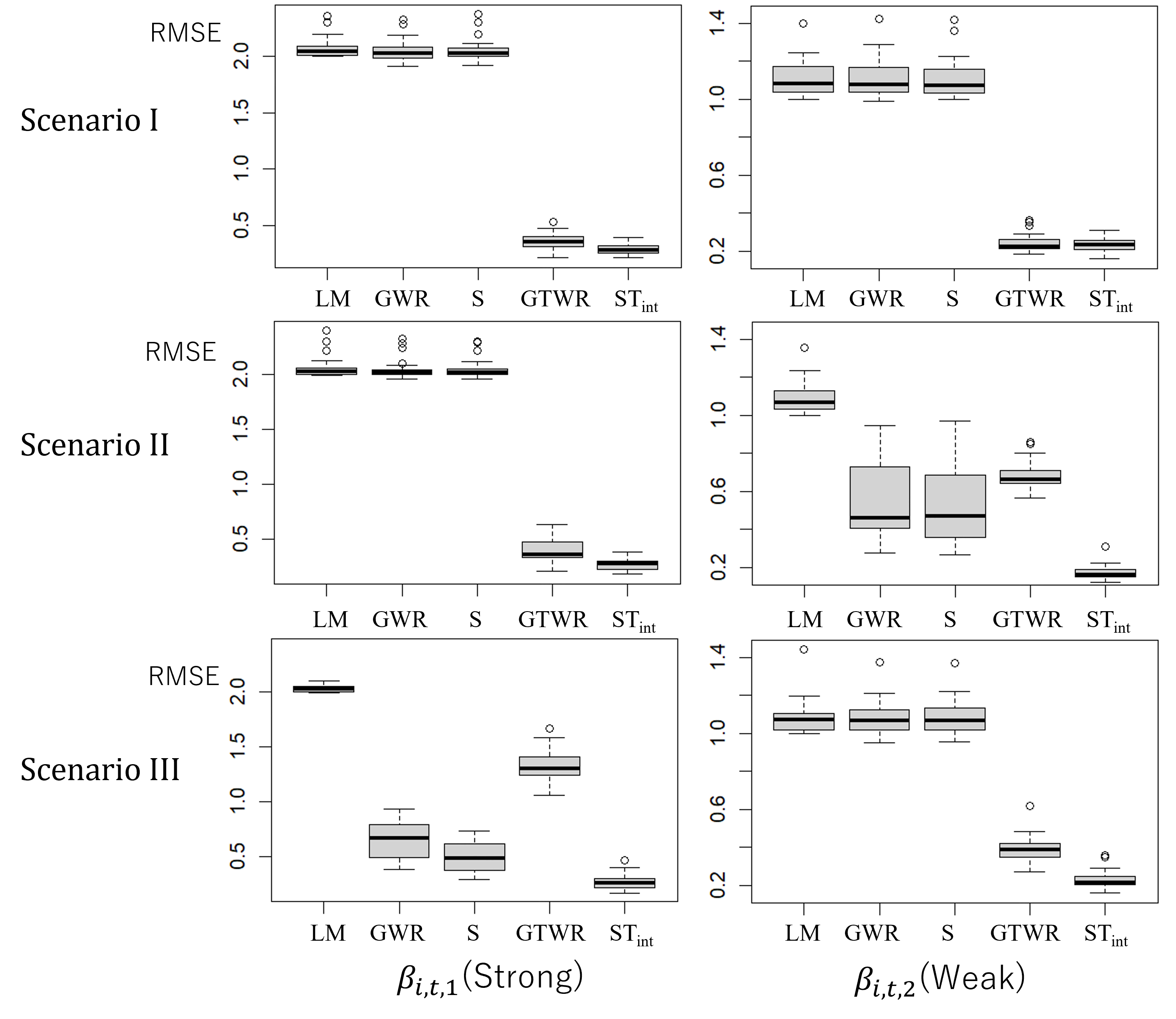}
    \caption{Boxplots of the RMSEs for the regression coefficients.}
    \label{fig_RMSE_GWR}
\end{figure}

Figure\ref{fig_RMSE_GWR} compares the RMSEs for the strong and weak STVCs. For reference, LM and S are compared again. In scenario I, GTWR and $\rm{STc_{int}}$ outperform GWR and S, confirming the importance of considering temporal patterns in regression coefficients. However, GTWR exhibits higher RMSEs compared to the spatial models for the SVCs assumed in $\beta_3(s,t)$ in scenario II and $\beta_2(s,t)$ in scenario III. This trend is pronounced in scenario III, where a larger variation is assumed in the SVC. In contrast, our $\rm{STc_{int}}$ method demonstrates better accuracy than both S and GWR, even when SVC is the true process. These findings suggest that our method is more accurate and robust to the underlying spatio-temporal structure of the coefficients compared to the alternative methods. The superior performance of $\rm{STc_{int}}$ is attributed to its flexibility in accommodating the differences in smoothness and variation of each spatial and/or temporal process in each coefficient, whereas GTWR assumes uniform bandwidth/scale parameters across the coefficients.

The median computation times in scenario I are 0.905 (GWR), 0.993 (S), 110.2 (GTWR), and 1.792 seconds, respectively. Owing to the proposed algorithm, $\rm{STc_{int}}$ is significantly faster and more accurate alternative to GTWR.


\section{Application}\label{sec_app1}

\subsection{Outline}
\begin{figure}[hbtp]
 \centering
 \includegraphics[scale=0.57]{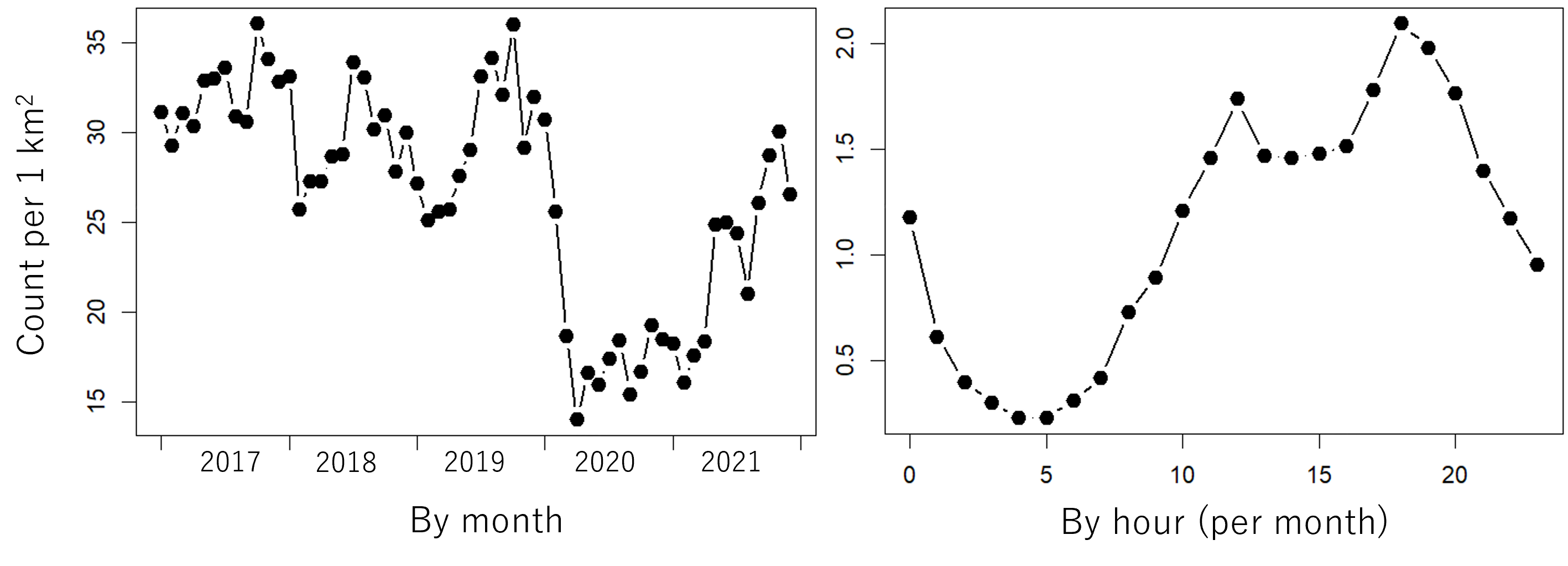}
 \caption{Larceny density by month (left) and hour (right)}
 \label{fig_larecny_time}
\end{figure}

\begin{figure}[hbtp]
 \centering
 \includegraphics[scale=0.55]{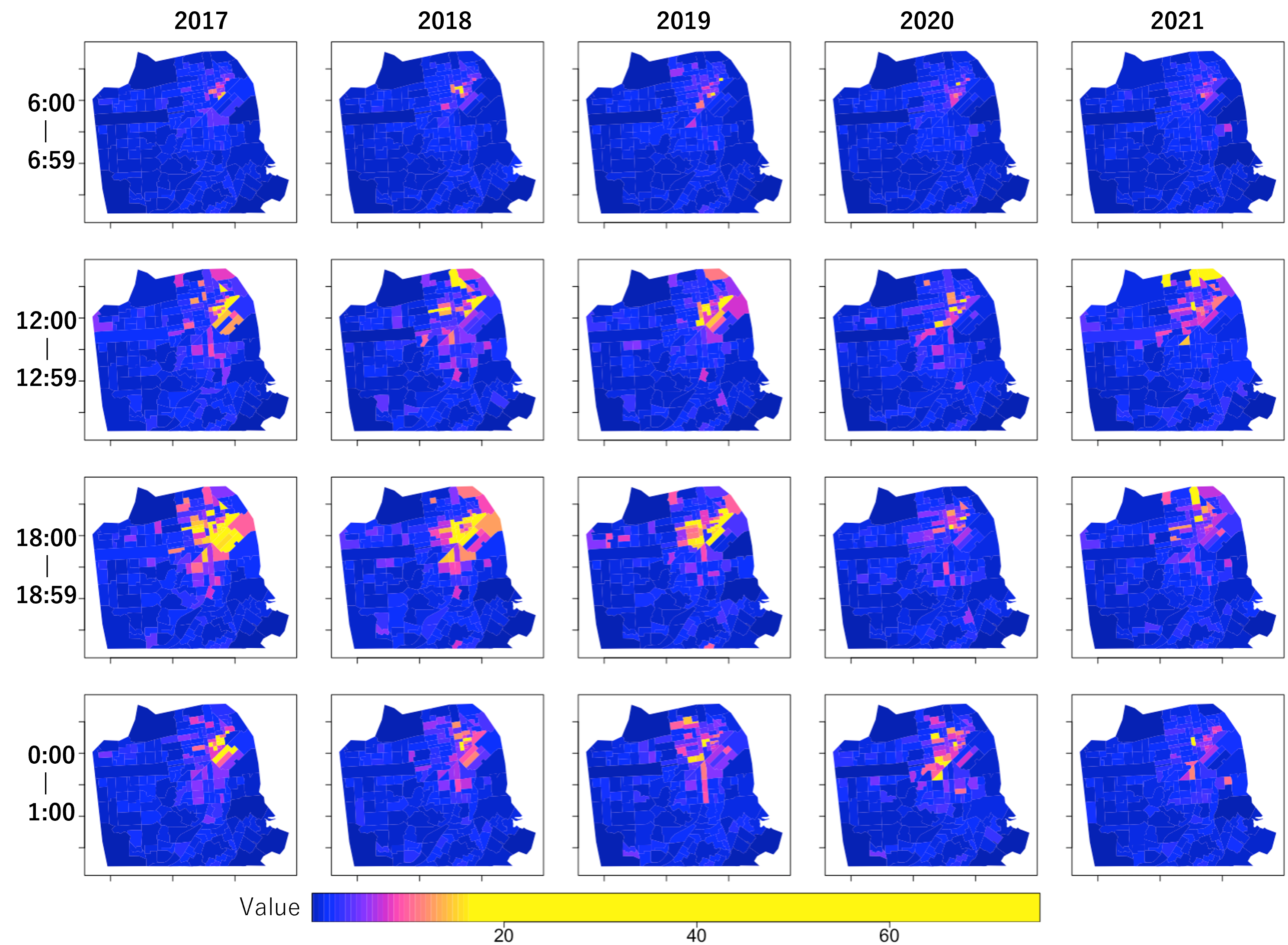}
 \caption{Larceny density in April by district by year by hour}
 \label{fig_larecny_space}
\end{figure}

This section applies the proposed method to analyze hourly and monthly larceny counts per  square km, which we will call the larceny density (source: Crime Dashboard: \url{https://www.sanfranciscopolice.org/stay-safe/crime-data/crime-dashboard}), by 194 districts between 2017 and 2021 ($N=279,360$) in San Francisco. As shown in Fig. \ref{fig_larecny_time} (left),  larceny drastically decreased in early 2020 due to the coronavirus disease (COVID-19), and gradually recovered thereafter. A cyclic pattern, increasing during the day and decreasing at night, is also evident from Fig. \ref{fig_larecny_time} (right).  Larceny densities mapped in Fig. \ref{fig_larecny_space} show that larceny was concentrated in the northeastern city center, especially during 2017 - 2019, which was before COVID-19, during the daytime hours.


\begin{figure}[hbtp]
 \centering
 \includegraphics[scale=0.35]{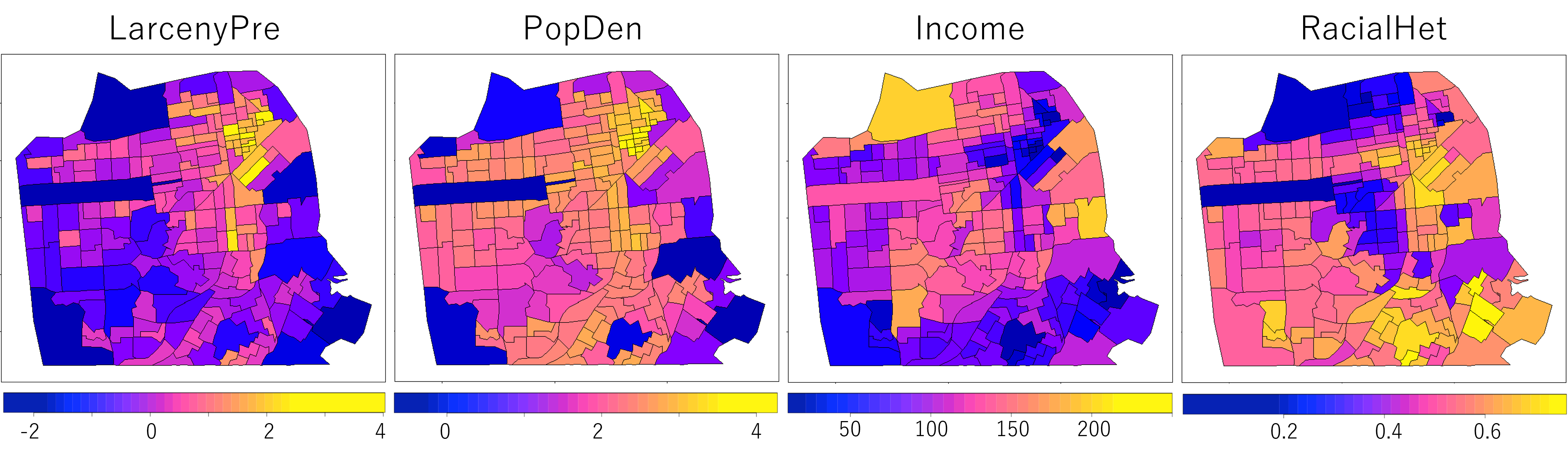}
 \caption{Covariates in April 2019}
 \label{fig_covariates}
\end{figure}

The explained variable is the logged larceny density. The covariates include logged larceny density in the previous month (LarenyPre), logged population density (PopDen; 1,000 people; yearly), median income (Income; 1,000 USD; yearly), and racial heterogeneity (RacialHet; yearly) measured by $1-\Sigma_{i=1}^7 a^2_i$, where $a_i$ is the share of the $i$-th racial population among the seven racial groups defined by the U.S. census (Fig.\ref{fig_covariates}). LarenyPre is considered to estimate the strength of the repetitive tendency of crime at a neighborhood-level; this tendency, which is called near repeat victimization \cite{amemiya2020near}, is attributable to the groups committing crimes in the same area or spatial concentration of facilities with risk.\footnote{ LarenyPre is a temporal lag of the explained variable. Although the least squares estimator of the coefficient is biased, it is consistent under standard assumptions (e.g., \cite{brockwell1991time}), meaning that the bias asymptotically disappears for large samples like our case. Besides, if explained variables are normally distributed, as we assumed, the estimator has asymptotic optimal properties \cite{lutkepohl2004applied}. Therefore, LarcenyPre is treated same as other covariates, and the coefficient is estimated using the penalized least squares estimator (Eq. \ref{eq:penPLS}).} RacialHet is considered because a higher racial heterogeneity can lower the level of social control \cite{cahill2007using}. These variables are sourced from the American Community Survey Tables (\url{https://www.census.gov/programs-surveys/acs/}). 

The six models $\rm{LM, S, ST, ST_{int}, STc, STc_{int}}$ considered in Section \ref{sec_mc} were compared. Here, $\rm{ST, ST_{int}}$, and $\rm{STc_{int}}$ consider non-cyclic time series by month during 2017 - 2021 while STc and $\rm{STc_{int}}$ consider cyclic time series by hour. They are estimated without parallelization.

\subsection{Result}
Table \ref{e_stat} summarizes the error statistics and computation time. LM achieves a reasonable accuracy because it implicitly considers dynamic spatio-temporal patterns through LarcenyPre. However, S demonstrates superior adjusted R-squares ($R_{adj}^2$), log-likelihood, and BIC values, indicating that SVC greatly improves the modeling accuracy. The accuracy is further improved by considering temporal patterns (ST and $\rm{ST_{int}}$) or space-time interactions ($\rm{ST_{int}}$ and $\rm{STc_{int}}$) in regression coefficients. $\rm{STc_{int}}$, which considers all of these, achieved the best accuracy.

Regarding computation time, $\rm{STc_{int}}$ tends to be slower than the other simpler models, as expected. However, it took only 832 seconds for the large samples with $N=279,360$, confirming the computational efficiency of the proposed method.

\begin{table}[h]
 \caption{Error statistics and computation time (CP time) in seconds. $R_{adj}^2$ denotes (conditional) adjusted R-squares}\label{e_stat}
 \centering
\begin{tabular}{lrrrrrr}
\hline
 & LM & S & ST & $\rm{ST_{int}}$ & STc & $\rm{STc_{int}}$ \\
 \hline
$R_{adj}^2$ & 0.666&0.722&0.726&0.727&0.740&0.745 \\
Log-likelihood& -240,576&-215,552&-214,046&-213,589&-206,685&-204,235 \\
BIC & 481,228&431,305&428,417&427,503&413,822&408,922 \\
\hline
CP time & 0.052 & 40.489& 188.351& 722.452& 303.724&832.219\\
\hline
\end{tabular}
\end{table}

Figure\ref{anova_xb} compares the variances explained by each explanatory variable in $\rm{STc_{int}}$. LarcenyPre and PopDen account for a large amount of variation, while Income and RacialHet do not contribute substantially. This suggests that larceny tends to occur repeatedly in the same neighborhoods, particularly in the densely populated districts.

\begin{table}[h]
 \caption{Variance explained by each regression term $x_p(s,t)\hat{\beta}_p(s,t)$ in $\rm{STc_{int}}$.}\label{anova_xb}
 \centering
\begin{tabular}{ccccc}
\hline
Intercept&LarcenyPre&PopDen&Income&RacialHet\\
 \hline
0.114 & 0.146 & 0.107 & 0.024 & 0.015\\\hline
\end{tabular}
\end{table}

\begin{table}[h]
 \caption{Proportion of the variances in the varying coefficients estimated by $\rm{STc_{int}}$. Zeros, representing the unselected variables, are omitted for simplicity. }\label{anova}
 \centering
\begin{tabular}{lccccc}
\hline
&Intercept&LarcenyPre&PopDen&Income&RacialHet\\
 \hline
Spatial& 0.922 & 0.624 & 0.854 & 0.958 & 0.983\\
Month& 0.008 &0.284&0.051&0.013&0.016\\
Hour& 0.048 &0.092&0.047&0.029&0.001 \\
Spatial$\times$Month& & & & & \\
Spatial$\times$Hour&0.022& &0.048& & \\\hline
\end{tabular}
\end{table}

Table \ref{anova} summarizes the proportion of variances in each STVC. This table suggests that the determinants of larceny risk are spatially varying but less time varying. Exceptionally, the coefficients on LarcenyPre and PopDen have relatively strong monthly and hourly temporal variations. As explained later, the former strongly reflects COVID-19, while the latter captures the periodic pattern in daily urban activity. In particular, 28.4 \% of the variation in the LarcenyPre coefficient is attributed to the month, while 9.2 \% is attributed to the hour. This suggests that LarcenyPre exhibits a strong seasonal pattern and a weak hourly pattern. 

Space-time interaction is found only in the spatially varying intercept and the coefficient on PopDen. Non-interaction terms were sufficient to capture most coefficient variations. If all interaction and non-interaction terms were equally considered, as in the existing studies, the number of basis functions would be very large, and the resulting model would be complex and computationally inefficient. The utility of the proposed method is confirmed in identifying an accurate and parsimonious STVC model computational efficiency.

\begin{figure}[hbtp]
 \centering
 \includegraphics[scale=0.82]{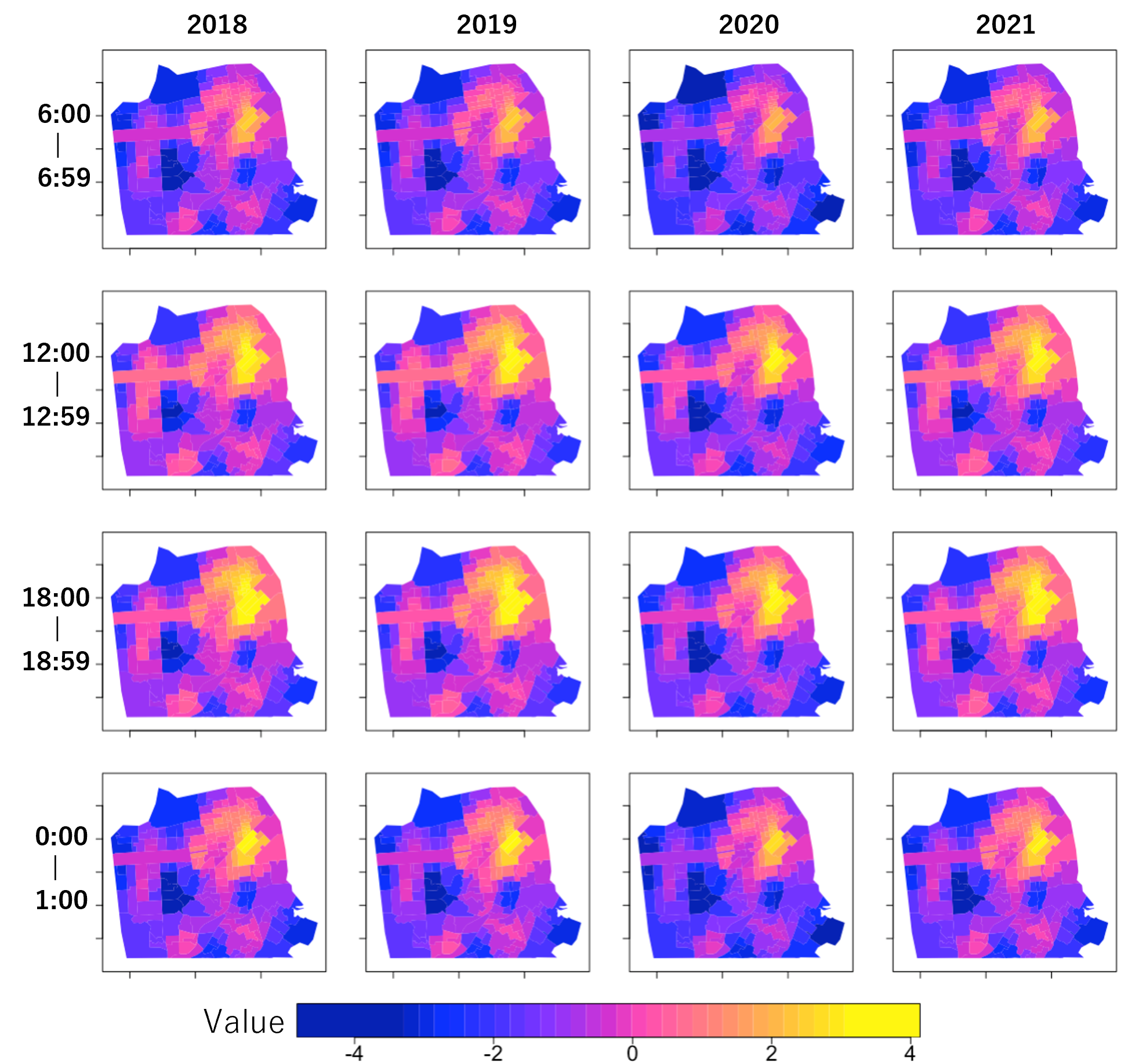}
 \caption{Estimated varying intercept in April by month by hour}
 \label{fig_intercept}
\end{figure}

\begin{figure}[hbtp]
 \centering
 \includegraphics[scale=0.82]{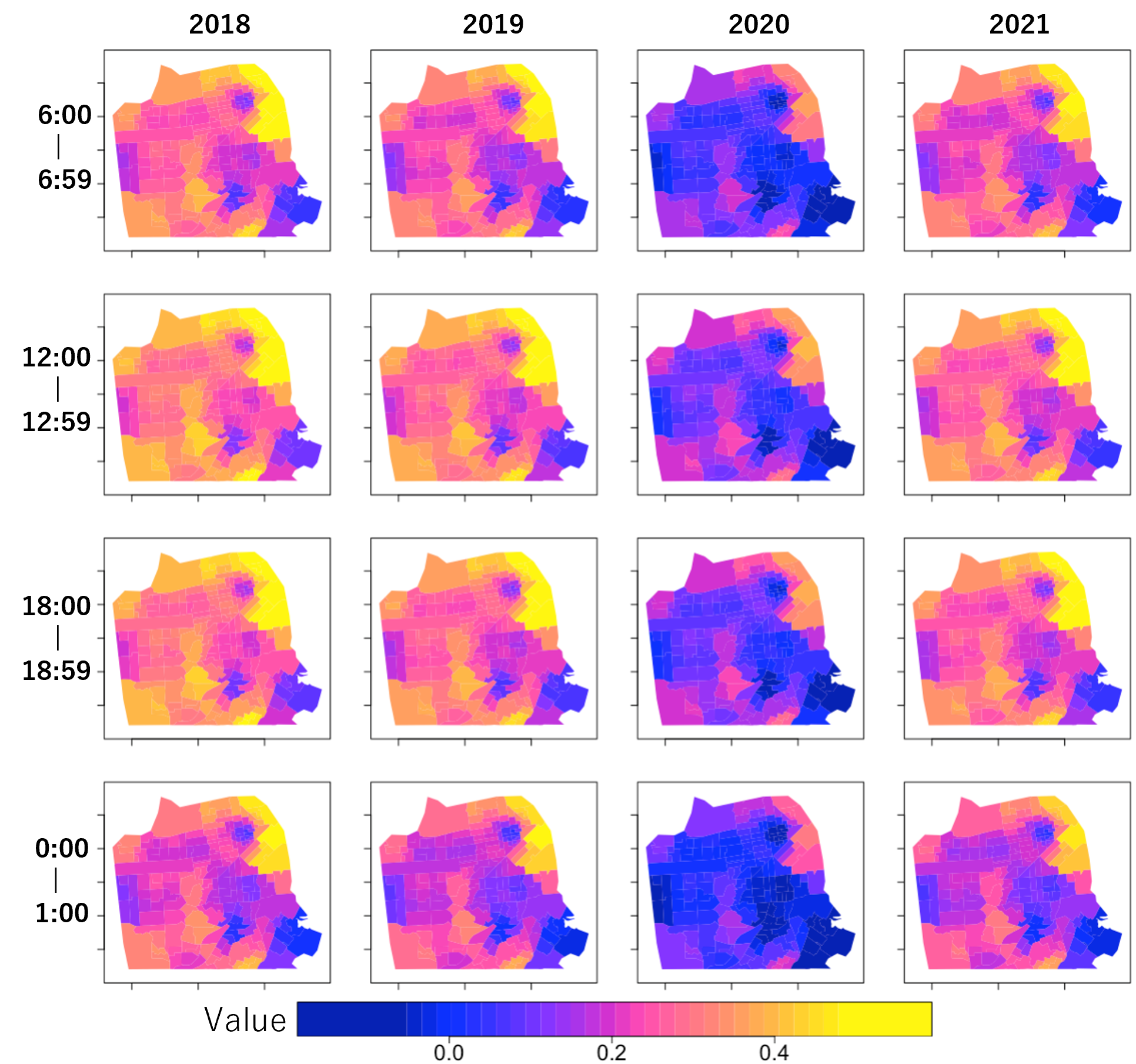}
 \caption{Estimated varying coefficient of LarcenyPre in April by month by hour}
 \label{fig_larcenypre}
\end{figure}

\begin{figure}[hbtp]
 \centering
 \includegraphics[scale=0.82]{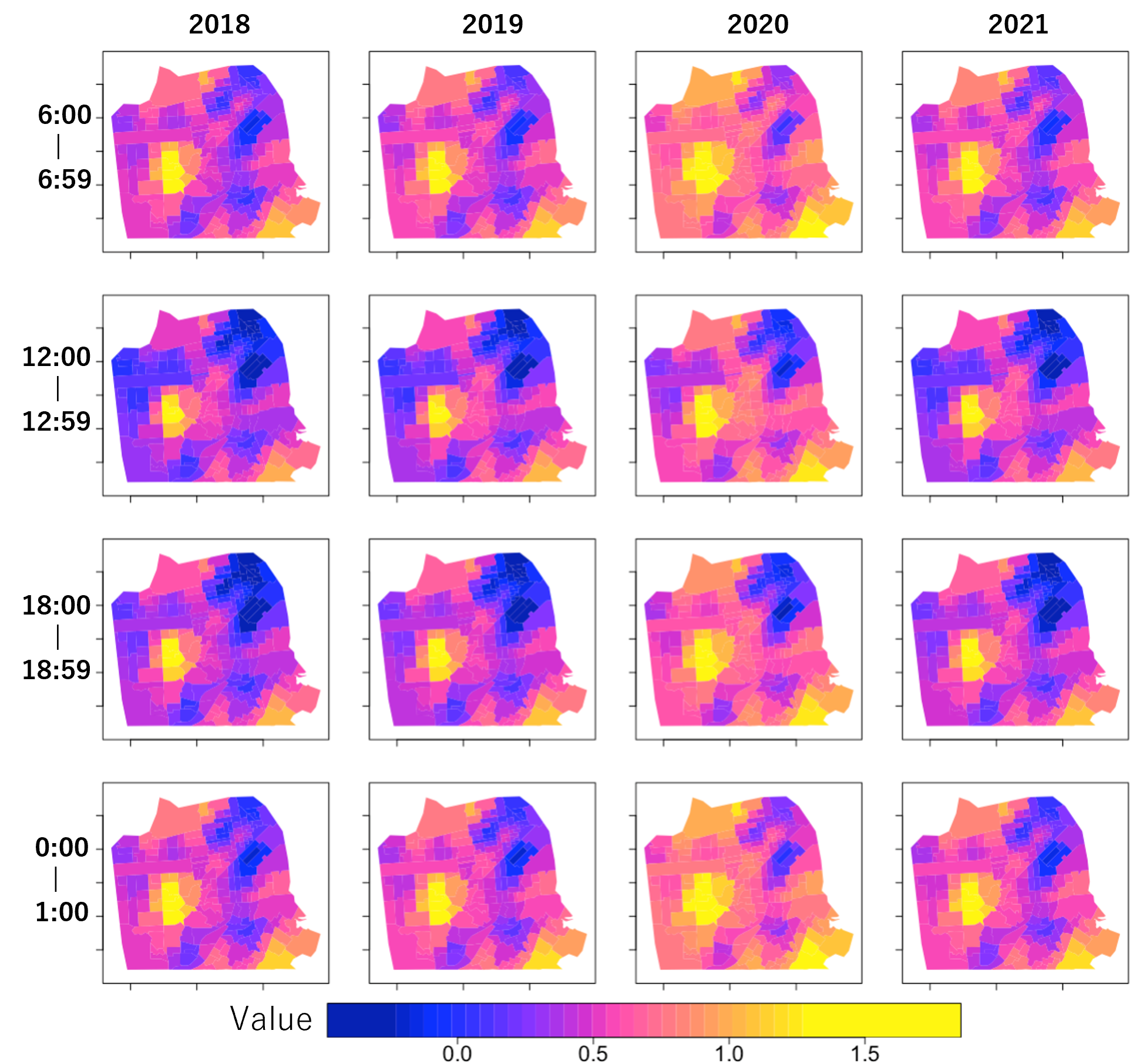}
 \caption{Estimated varying coefficient of PopDen in April by month by hour}
 \label{fig_popden}
\end{figure}

Figures \ref{fig_intercept} - \ref{fig_popden} map the estimated varying intercept and STVCs for LarcenyPre and PopDen, which exhibit relatively strong time variations. Based on the intercept, the northeastern city center has been the hot-spot over the years, and this tendency becomes more pronounced during the day. As expected, this map pattern is consistent with the observed larceny distribution in Fig. \ref{fig_larecny_space}.

Before 2020, coefficients on LarcenyPre were positive across areas, with particularly high values observed in the northeastern bayside area. This suggests a pattern of repeated larceny occurrences within the same neighborhood, especially in the bayside area. However, this repetitive tendency significantly weakened in 2020 with the onset of COVID-19. Based on the result, larceny occurrences became randomized and hard to predict during the pandemic. By 2021, the repetitive tendency recovered to the before COVID-19 level. It is also conceivable that larceny was more likely to recur within the same area during the day than at night. 

The coefficients of PopDen consistently exhibit high values in a middle western area. According to Fig. \ref{fig_covariates}, this hotspot has a relatively high income level compared to the neighboring areas, suggesting a higher risk in the local populated and affluent area. In contrast, coefficients of PopDen decline in the northeastern central area where income levels are lower, indicating that the risk of larceny in populated areas may depend on the neighborhood wealth. In 2020, coefficients of PopDen slightly increased relative to other years. This was due to the increased chance of committing a larceny near one's home rather than in the city center owing to the movement restrictions during the pandemic. The increase in coefficients was particularly noticeable at night, suggesting a higher risk at nighttime in populated areas during the pandemic year.

\begin{figure}[hbtp]
 \centering
 \includegraphics[scale=0.5]{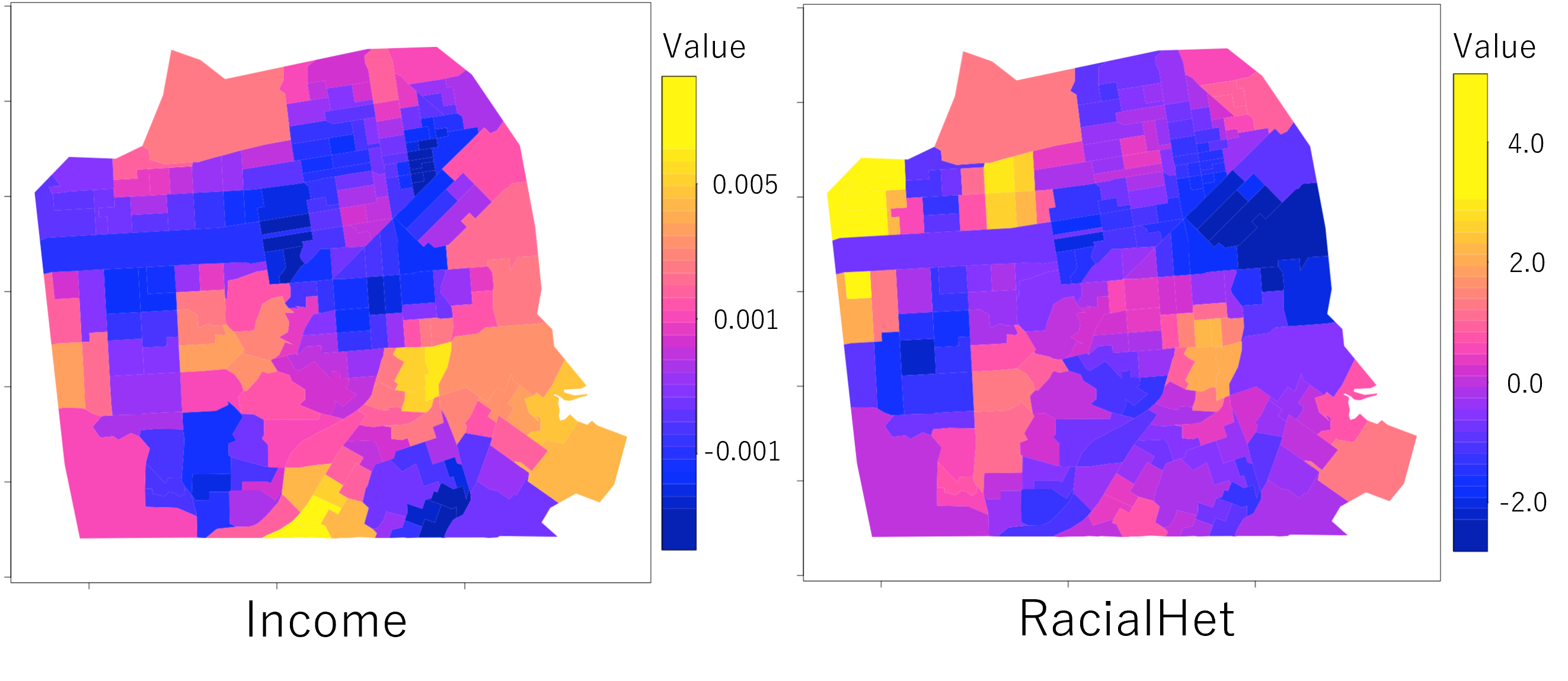}
 \caption{Estimated varying coefficients of other explanatory variables in April 2019 during 12:00 - 12:59. These coefficients are spatially varying but approximately constant over time.}
 \label{fig_other}
\end{figure}

The coefficients of Income and RacialHet, which are estimated to be nearly constant over time, are plotted for August 2019 in Fig. \ref{fig_other}. The coefficients on Income show large positive values in the southeastern area where income levels are low, suggesting that a higher income might increase the risk of larceny in this area. Although the coefficients of RacialHet tend to be high in several northwestern districts, their influence is marginal owing to their weak explanatory power, as shown in Table \ref{anova_xb}.

In summary, the proposed method effectively identified space-time patterns in multiple time scales within each coefficient in a computationally efficient manner. The modeling accuracy for crime was enhanced compared to the conventional spatial model (S), ignoring the temporal variation. Based on the empirical result, the determinants of larceny risk vary spatially but exhibit less temporal variation. This suggests that effective countermeasures need to be tailored to specific neighborhoods.

\section{Concluding remarks}\label{sec_remark}
This study develops a scalable STVC model and a procedure to select the optimal varying coefficient specification. Unlike existing methods, the proposed method explicitly distinguishes between the main/non-interaction and interaction effects. The resulting method accurately and computationally efficiently identifies multiple spatial and temporal processes within regression coefficients, as demonstrated in Monte Carlo experiments. Additionally, through crime modeling, it is also confirmed that the proposed method provides intuitively reasonable coefficient estimates. 

Cyclic temporal patterns by hour, day of the week, and season are ubiquitous as well as dynamic temporal patterns. As demonstrated in Section \ref{sec_app1}, the strength of some regression coefficients exhibits cyclic patterns. Nevertheless, cyclic patterns have rarely been considered in STVC modeling. Our empirical study highlights the importance of considering not just spatial but also cyclic and non-cyclic temporal (and space-time) patterns in regression coefficients.

However, some challenges remain to be resolved. Specifically, the developed method assumes a Gaussian explained variable. Algorithm must be extended to handle counts, binary, or other non-Gaussian data. An extension under the generalized linear model framework appears promising to overcome this limitation, although a time consuming iteration of the weighed least squares estimation is likely needed. Alternatively, transforming the explained variables is another approach that potentially avoids such an additional iteration. In fact, a Gaussian SVC model has been extended to various non-Gaussian models by introducing a flexible transformation function, as shown in \cite{murakami2021compositionally}. A similar transformation approach is also available to approximate a Poisson model for count data with accuracy comparable to the usual generalized linear model \cite{murakami2022improved}. The introduction of these transformation functions would enhance the applicability of our method for a wide variety of real-world problems.

The proposed STVC model is implemented in the \verb|resf_vc| function in an R package spmoran (\url{https://cran.r-project.org/web/packages/spmoran/index.html}). See a support site (\url{https://github.com/dmuraka/spmoran}) for details on how to implement this model.

\begin{appendices}
\section{Appendix: Derivation of the marginal likelihood} \label{app_lik}

Let us express our model (Eq. \ref{eq_mat}) as follows:
\begin{equation}
    \bm{y}=\bm{Xb}+\bm{Z_K}\bm{V}(\bm{\Theta_K})^{1/2}\bm{u_K}+\bm{\epsilon}, \hspace{0.5cm}
    \bm{u_K}\sim N(\bm{0},\sigma^2\bm{I}),\hspace{0.5cm}
    \bm{\epsilon}\sim N(\bm{0},\sigma^2\bm{I}),
\end{equation}
where $\bm{\gamma_K}=\bm{V}(\bm{\Theta_K})^{1/2}\bm{u_K}$. The model is expressed in distribution form as
\begin{equation}
    p(\bm{y},\bm{u_K}|\bm{b},\bm{\Theta_K},\sigma^2)=p(\bm{y}|\bm{b},\bm{u_K},\bm{\Theta_K},\sigma^2)p(\bm{u_K}|\sigma^2),
\end{equation}
where
\begin{align}
p(\bm{y}|\bm{b},\bm{u_K},\bm{\Theta_K},\sigma^2)=&\frac{1}{(2\pi\sigma^2)^{N/2}} \exp \left(-\frac{\bm{\epsilon}'\bm{\epsilon}}{2\sigma^2} \right),\label{eq_y_dist1} \\
p(\bm{u_K}|\bm{\Theta_K},\sigma^2)=&\frac{1}{(2\pi\sigma^2)^{\tilde{L}/2}} \exp \left(-\frac{\bm{u_K}'\bm{u_K}}{2\sigma^2} \right).\label{eq_y_dist2}
\end{align}
$\tilde{L}$ represents the dimension of $\bm{u_K}$. By multiplying Eqs. (\ref{eq_y_dist1} - \ref{eq_y_dist2}), we have
\begin{equation}
    p(\bm{y},\bm{u_K}|\bm{b},\bm{\Theta_K},\sigma^2)=\frac{1}{(2\pi\sigma^2)^{(N+\tilde{L})/2}}\exp\left(-\frac{\bm{\epsilon}'\bm{\epsilon} + \bm{u_K}'\bm{u_K} }{2\sigma^2} \right). \label{eq_joint_dist}
\end{equation}
By applying the Taylor expansion around an estimate $\bm{\hat{c}_K}=[\hat{\bm{b}}',\hat{\bm{u}}_{\bm{K}}']'$ of $\bm{c_K}=[\bm{b}',\bm{u_K}']'$, we obtain
\begin{equation}
\begin{split}    
    \log p(\bm{y},\bm{u_K}|\bm{b},\bm{\Theta_K},\sigma^2)=&\log p(\bm{y},\hat{\bm{u}}_{\bm{K}}|\hat{\bm{b}},\bm{\Theta_K},\sigma^2) + \\
    &\frac{1}{2}(\bm{c_K}-\hat{\bm{c}}_{\bm{K}})' 
    \frac{ \partial^2 \log p(\bm{y},\bm{u_K}|\bm{b},\bm{\Theta_K},\sigma^2) }{ \partial \bm{c_K} \partial \bm{c}'_{\bm{K}} }
    (\bm{c_K}-\hat{\bm{c}}_{\bm{K}}),
\end{split}
\end{equation}
where no reminder terms exist because higher order terms are zero since $\bm{y}$ and $\bm{c_K}$ are Gaussians. The Hessian matrix equals $\frac{ \partial^2 \log p(\bm{y},\bm{u_K}|\bm{b},\bm{\Theta_K}) }{ \partial \bm{c_K} \partial \bm{c}'_{\bm{K}} }=\bm{H_{K,K}}$ \cite{wood2017generalized}. By organizing the equality, we have
\begin{equation}
\bm{\epsilon}'\bm{\epsilon} + \bm{u_K}'\bm{u_K} = \hat{\bm{\epsilon}}'\hat{\bm{\epsilon}} + \bm{\hat{u}_K}'\bm{\hat{u}_K} + 
(\bm{c_K}-\hat{\bm{c}}_{\bm{K}})'\bm{H_{K,K}}(\bm{c_K}-\hat{\bm{c}}_{\bm{K}}),
\end{equation}
where $\hat{\bm{\epsilon}}=\bm{y}-\bm{X\hat{b}}-\bm{Z_K}\bm{V}(\bm{\Theta_K})^{1/2}\hat{\bm{u}}_{\bm{K}}$. By substituting it into Eq. (\ref{eq_joint_dist}), we have
\begin{equation}
\begin{split}
p(\bm{y},\bm{u_K}|\bm{b},\bm{\Theta_K},\sigma^2)=&\frac{1}{(2\pi\sigma^2)^{(N+\tilde{L})/2}}\exp\left(-\frac{\hat{\bm{\epsilon}}'\hat{\bm{\epsilon}} + \bm{\hat{u}_K}'\bm{\hat{u}_K} + 
(\bm{c_K}-\hat{\bm{c}}_{\bm{K}})'\bm{H_{K,K}}(\bm{c_K}-\hat{\bm{c}}_{\bm{K}}) }{2\sigma^2} \right)\\
=&\frac{(2\pi\sigma^2)^{(P+\tilde{L})/2}|\bm{H_{K,K}}|^{-1/2}}{(2\pi\sigma^2)^{(N-P)/2}}\exp\left(-\frac{\hat{\bm{\epsilon}}'\hat{\bm{\epsilon}} + \bm{\hat{c}_K}'\bm{\hat{c}_K} }{2\sigma^2} \right) \times \\
&\frac{1}{(2\pi\sigma^2)^{(P+\tilde{L})/2}|\bm{H_{K,K}}|^{-1/2}} \exp \left( \frac{-(\bm{c_K}-\hat{\bm{c}}_{\bm{K}})'\bm{H_{K,K}}(\bm{c_K}-\hat{\bm{c}}_{\bm{K}}) }{2\sigma^2} \right).
\end{split}\label{eq_marg0}
\end{equation}
Because the third line represents the probability density function of $\bm{c_K}\sim N(\hat{\bm{c}}_{\bm{K}}, \sigma^2\bm{H_{K,K}}^{-1})$,
\begin{equation}
    \int \frac{1}{(2\pi\sigma^2)^{(P+\tilde{L})/2}|\bm{H_{K,K}}|^{-1/2}} \exp \left( \frac{-(\bm{c_K}-\hat{\bm{c}}_{\bm{K}})'\bm{H_{K,K}}(\bm{c_K}-\hat{\bm{c}}_{\bm{K}}) }{2\sigma^2} \right) d \bm{c_K} = 1,\label{eq_marg}
\end{equation}
Let us evaluate $ p(\bm{y}|\bm{\Theta_K},\sigma^2)=\int p(\bm{y},\bm{u_K}|\bm{b},\bm{\Theta_K},\sigma^2)d \bm{c_K}$ using Eq. (\ref{eq_marg}). Subsequently, we obtain
\begin{equation}
p(\bm{y}|\bm{\Theta_K},\sigma^2)=\frac{|\bm{H_{K,K}}|^{-1/2}}{(2\pi\sigma^2)^{(N-P)/2}}\exp\left(-\frac{\hat{\bm{\epsilon}}'\hat{\bm{\epsilon}} + \bm{\hat{u}_K}'\bm{\hat{u}_K} }{2\sigma^2} \right).
\end{equation}
The maximum likelihood estimate $\hat{\sigma}^2=\frac{\hat{\bm{\epsilon}}'\hat{\bm{\epsilon}} + \bm{\hat{u}_K}'\bm{\hat{u}_K}}{N-P}$ of $\sigma^2$ is readily obtained by solving $\frac{\partial p(\bm{y}|\bm{\Theta_K},\sigma^2)}{\partial \sigma^2}=0$. By substituting the estimate, the likelihood becomes
\begin{equation}
p(\bm{y}|\bm{\Theta_K})=\frac{|\bm{H_{K,K}}|^{-1/2}}{\left(2\pi\frac{\hat{\bm{\epsilon}}'\hat{\bm{\epsilon}} + \bm{\hat{u}_K}'\bm{\hat{u}_K}}{N-P}\right)^{(N-P)/2}}\exp\left(-\frac{N-P}{2} \right).
\end{equation}
By taking the logarithm, our marginal log-likelihood (Eq. \ref{eq_loglik}) is derived.

\end{appendices}

\bibliographystyle{abbrv}
\bibliography{refs}
\end{document}